\documentclass[useAMS,usenatbib,preprint]{aastex}
\usepackage{txfonts,graphicx,natbib,color,verbatim,footmisc}
\bibpunct{(}{)}{;}{a}{}{,}

\def \apjs {ApJS}
\def \apjl {ApJL}
\def \aap {A\&A}

\def\lesssim{\mathrel{\hbox{\rlap{\hbox{\lower4pt\hbox{$\sim$}}}\hbox{$<$}}}}
\def\gtrsim{\mathrel{\hbox{\rlap{\hbox{\lower4pt\hbox{$\sim$}}}\hbox{$>$}}}}

\DeclareMathAlphabet{\mathsc}{OT1}{cmr}{m}{sc}
\def\testbx{bx}%
\DeclareRobustCommand{\ion}[2]{%
\relax\ifmmode
\ifx\testbx\f@series
{\mathbf{#1\,\mathsc{#2}}}\else
{\mathrm{#1\,\mathsc{#2}}}\fi
\else\textup{#1\,{\mdseries\textsc{#2}}}%
\fi}

\begin{document}
\label{firstpage}

 \title{Anatomy of a post-starburst minor merger: a multi-wavelength WFC3 study of NGC 4150\footnotemark}
 \shorttitle{Recent star formation in NGC 4150}
 \shortauthors{Crockett et al.}

\footnotetext[1]{Based on observations with the NASA/ESA Hubble Space Telescope obtained at the Space Telescope Science Institute, which is operated by the Association of Universities for Research in Astronomy, Incorporated, under NASA contract NAS5-26555.}

%%% WFC3 SOC AUTHOR LIST FOR VERS PAPERS: %%% Luppino, MacKenty prefer not to be included
\author{
R. Mark Crockett,\altaffilmark{2} 
Sugata Kaviraj,\altaffilmark{2,3} 
Joseph I.~Silk,\altaffilmark{2}
Bradley C.~Whitmore,\altaffilmark{4} 
Robert W.~O'Connell,\altaffilmark{5}
Max Mutchler,\altaffilmark{4} 
Bruce Balick,\altaffilmark{6} 
Howard E.~Bond,\altaffilmark{4} 
Daniela Calzetti,\altaffilmark{7} 
C. Marcella Carollo,\altaffilmark{8} 
Michael J.~Disney,\altaffilmark{9} 
Michael A.~Dopita,\altaffilmark{10} 
Jay A.~Frogel,\altaffilmark{11} 
Donald N.~B.~Hall,\altaffilmark{12}
Jon A.~Holtzman,\altaffilmark{13} 
Randy A.~Kimble,\altaffilmark{14} 
Patrick J.~McCarthy,\altaffilmark{15}
Francesco Paresce,\altaffilmark{16} 
Abhijit Saha,\altaffilmark{17}
John T.~Trauger,\altaffilmark{18} 
Alistair R.~Walker,\altaffilmark{19} 
Rogier A.~Windhorst,\altaffilmark{20}
Erick T.~Young,\altaffilmark{21} 
Hyunjin Jeong\altaffilmark{22} 
and 
Sukyoung K.~Yi\altaffilmark{22}}

\altaffiltext{2}{Department of Physics, University of Oxford, Oxford OX1 3RH, United Kingdom; mark.crockett@astro.ox.ac.uk}
\altaffiltext{3}{Astrophysics Group, Imperial College London, South Kensington Campus, London SW7 2AZ, United Kingdom}
\altaffiltext{4}{Space Telescope Science Institute, Baltimore, MD 21218} 
\altaffiltext{5}{Department of Astronomy, University of Virginia, Charlottesville, VA 22904-4325}
\altaffiltext{6}{Department of Astronomy, University of Washington, Seattle, WA 98195-1580} 
\altaffiltext{7}{Department of Astronomy, University of Massachusetts, Amherst, MA 01003}
\altaffiltext{8}{Department of Physics, ETH-Zurich, Zurich, 8093 Switzerland} 
\altaffiltext{9}{School of Physics and Astronomy, Cardiff University, Cardiff CF24 3AA, United Kingdom}
\altaffiltext{10}{Research School of Astronomy \& Astrophysics, The Australian National University, ACT 2611, Australia}
\altaffiltext{11}{Association of Universities for Research in Astronomy, Washington, DC 20005} 
\altaffiltext{12}{Institute for Astronomy, University of Hawaii, Honolulu, HI 96822}
\altaffiltext{13}{Department of Astronomy, New Mexico State University, Las Cruces, NM 88003} 
\altaffiltext{14}{NASA--Goddard Space Flight Center, Greenbelt, MD 20771}
\altaffiltext{15}{Observatories of the Carnegie Institution of Washington, Pasadena, CA 91101-1292} 
\altaffiltext{16}{Istituto di Astrofisica Spaziale e Fisica Cosmica, INAF, Via Gobetti 101, 40129 Bologna, Italy} 
\altaffiltext{17}{National Optical Astronomy Observatories, Tucson, AZ 85726-6732} 
\altaffiltext{18}{NASA--Jet Propulsion Laboratory, Pasadena, CA 91109} 
\altaffiltext{19}{Cerro Tololo Inter-American Observatory, La Serena, Chile}
\altaffiltext{20}{School of Earth and Space Exploration, Arizona State University, Tempe, AZ 85287-1404}
\altaffiltext{21}{NASA--Ames Research Center, Moffett Field, CA 94035} 
\altaffiltext{22}{Department of Astronomy, Yonsei University, Seoul 120-749, Korea}

\begin{abstract}
We present a spatially-resolved near-UV/optical study, using the
Wide Field Camera 3 (WFC3) on board the {\em Hubble Space
Telescope}, of NGC 4150; a sub-L$_{*}$, early-type galaxy (ETG) of
around $6\times10^9$ M$_{\odot}$, which has been observed as part
of the WFC3 Early-Release Science Programme.  Previous work
indicates that this galaxy has a large reservoir of molecular
hydrogen gas, exhibits a kinematically decoupled core (likely
indication of recent merging) and strong, central H$\beta$
absorption (indicative of young stars).  While relatively
uninspiring in its optical image, the core of NGC 4150 shows
ubiquitous near-UV emission and remarkable dusty substructure. Our
analysis shows this galaxy to lie in the near-UV {\em green
valley}, and its pixel-by-pixel photometry exhibits a narrow range
of near-UV/optical colours that are similar to those of nearby E+A
(post-starburst) galaxies, and lie between those of M83 (an
actively star-forming spiral) and the local quiescent ETG
population.  We parametrise the properties of the recent star
formation (age, mass fraction, metallicity and internal dust
content) in the NGC 4150 pixels by comparing the observed
near-UV/optical photometry to stellar models.  The typical age of
the recent star formation (RSF) is around 0.9 Gyrs, consistent
with the similarity of the near-UV colours to post-starburst
systems, while the morphological structure of the young component
supports the proposed merger scenario.  The typical RSF
metallicity, representative of the metallicity of the gas fuelling
star formation, is $\sim$0.3 - 0.5 Z$_{\odot}$.  Assuming that
this galaxy is a merger and that the gas is sourced mainly from
the infalling companion, these metallicities plausibly indicate
the gas-phase metallicity (GPM) of the accreted satellite.
Comparison to the local mass-GPM relation suggests (crudely) that
the mass of the accreted system is $\sim3\times10^8$ M$_{\odot}$,
making NGC 4150 a 1:20 minor merger.  A summation of the pixel RSF
mass fractions indicates that the RSF contributes $\sim$ 2-3
percent of the stellar mass.  This work reaffirms our hypothesis
that minor mergers play a significant role in the evolution of
ETGs at late epochs.

\end{abstract}


\maketitle

\setcounter{footnote}{22}

\section{Introduction}

\setlength\parindent{4mm}

Over recent decades, a central topic in observational
astrophysics has been the star formation histories (SFHs) of
massive early-type (elliptical and lenticular) galaxies.
Early-types host more than 50 per cent of the stellar mass
density in the nearby universe and, given their dominance of the
galaxy census, it is important that we develop a thorough
understanding of their formation and evolution.  The broad-band
optical colours of massive early-type galaxies are red and show a
tight correlation with luminosity; more luminous galaxies
possessing redder colours. Since early-types are not generally
dusty, their optical colours imply that the bulk of the
stellar population is old ($>$ 3 Gyrs) and perhaps coeval
\citep[e.g.,][]{1992MNRAS.254..589B,1998MNRAS.299.1193B,1997ApJ...483..582E,1998ApJ...492..461S,1998ApJ...501..571G,2000ApJ...541...95V}.
The tight fundamental plane and the high alpha-enhancement ratios
\citep[e.g.][]{1992ApJ...398...69W,1993MNRAS.265..553C,1997A&A...320...41K,2000AJ....120..165T}
further suggest small ranges in the epoch and duration of ETG star
formation.  However, a significant drawback of optical colours is
their insensitivity to moderate amounts of recent star formation
(RSF). Indeed, high precision optical spectroscopy has
consistently suggested extended periods of star formation lasting
until a few Gyr ago in some ETGs and has recently demonstrated
that the luminosity-weighted ages of local ETGs have a
large range and correlate with both velocity dispersion and
environment
\citep{2000AJ....120..165T,2005ApJ...632..137N,2005ApJ...621..673T,2009ApJ...698.1590G,2009MNRAS.398.1835S,2010MNRAS.404.1775T}.
The currently accepted Lambda Cold
Dark Matter ($\Lambda$CDM) paradigm suggests that this behavior is due to gas infall and
merger events that continue to recent times
\citep{2005MNRAS.360...60K}, although some of the cold gas fueling
star formation may be produced internally through stellar mass
loss \citep[e.g.][]{2006MNRAS.366.1151S,2005ApJ...634..258Y}.
Recent observational studies
\citep[e.g.][]{2010arXiv1001.2141K,2010arXiv1004.3775K} have
highlighted the role of {\em minor mergers} (progenitor
mass ratios $<$ 1:4) in the late-epoch ($z<1$) evolution of ETGs.

%The optical spectrum remains largely unaffected by the minority of stellar mass that is expected to form in these galaxies at low and intermediate redshift (z $<$ 1), which makes it difficult to quantify their recent SFHs using even the bluest optical filters (e.g. U; B). Indeed, until very recently, conclusive evidence of systematic recent star formation in early-type galaxies (which is predicted by the currently accepted Lambda Cold Dark Matter ($\Lambda$CDM) paradigm, due to e.g. minor mergers) remained elusive, and consensus strongly favoured models in which early-types were purely passively evolving systems with no measurable RSF at z $\sim$ 0.

An efficient way to pursue the problem of late star formation in
ETGs without the need for high S/N spectroscopy is to employ
rest-frame ultraviolet (UV; shortward of 3000 \AA) imaging. While
their impact on the optical spectrum is weak, small mass fractions
($<$ 5 per cent) of young ($<$ 0.5 Gyr old) stars can strongly
affect the near-UV (2000-3000\AA) (see Fig.~\ref{fig:col_vs_mass}).
Following the study by \citet{2005ApJ...619L.111Y},
\citet{2007ApJS..173..619K} comprehensively studied the UV
properties of a large sample ($\sim$2100) of massive (M$_{\rm  r}$
$<$ 21) early-type galaxies in the nearby Universe (0 $<$ z $<$
0.11), by combining UV data from the GALEX mission \citep{2005ApJ...619L...1M} with optical data from the Sloan Digital Sky
Survey \citep[SDSS;][]{2009ApJS..182..543A}.  When upper limits to the UV
flux from evolved stellar stages are considered, e.g. extreme
horizontal branch (EHB) stars which drive the `UV-upturn'
phenomenon typically observed in giant cluster ellipticals \citep{1999ApJ...513..128Y}, 20-30 per cent of massive early-type galaxies are
found to contain unambiguous signatures of RSF, because their
UV-optical SEDs cannot be produced by the EHB alone. It is worth
noting that since upper limits to the UV flux from old stars are
considered, the fraction of `star-forming' ETGs becomes a (robust)
lower limit. A more reliable result can be obtained by studying
the rest-frame UV properties of ETGs at $z>0.5$, where the
horizontal branch is not yet in place and the rest-frame UV
becomes a more reliable indicator of the presence of young stars.
By exploiting deep optical U and B-band data to trace the
rest-frame UV, \citet{2008MNRAS.388...67K} have demonstrated that
at these redshifts a negligible fraction of ETGs are consistent
with purely passive ageing since high redshift, supporting the low
redshift GALEX results and suggesting that RSF is a widespread
phenomenon in the ETG population over at least the last 8 billion
years.  The mass fractions forming in the RSF events range from
fractions of a percent to a few percent, with luminosity-weighted
ages of $\sim$ 300 - 500 Myrs \citep{2007ApJS..173..619K}. It is
worth noting that the large spread ($\sim$ 5 mags) in the
NUV-optical colours cannot be reproduced by RSF driven solely by
internal stellar mass loss, pointing to an external origin for the
gas that fuels that star formation. This seems consistent with
recent work that suggests that the RSF is driven by minor mergers.

The {\em Hubble Space Telescope} (HST) program exploited in this
study extends and enhances these results by exploring the
\emph{spatial distribution} of young stars in early-type
galaxies, which promises crucial insights into the
processes that lead to star formation in early-type systems.  In
this paper we use high resolution NUV-optical photometry from the
HST Wide Field Camera 3 (WFC3) to accurately quantify the age,
metallicity, mass and spatial distribution of young stars in the
early-type galaxy NGC 4150, with the aim of constraining the
characteristics of the suspected merger event (e.g. age,
mass-ratio) believed to have triggered the recent starburst.

\subsection{The target galaxy: NGC 4150}
\label{sec:target_n4150}

The properties of NGC 4150 are detailed in
Table~\ref{tab:ngc4150_props}.  This galaxy was also one of the
targets for the SAURON project
\citep{2001MNRAS.326...23B,2002MNRAS.329..513D}.  NGC 4150 is
characterised as a lenticular (S0) galaxy at
$\mathrm{\alpha_{J2000}=12^{h}10^{m}33^{s}.67,\delta_{J2000}=+30\degr24\arcmin05\arcsec.9}$,
which shows signs of RSF.Ê Previous observations have shown it to
possess blue NUV-optical colours \citep{2009MNRAS.398.2028J},
enhanced central H$\beta$ absorption \citep{2006MNRAS.369..497K},
large quantities of molecular gas \citep{2007MNRAS.377.1795C}, and
a kinematically decoupled core \citep{2008MNRAS.390...93K}.Ê The
first two characteristics are indicative of RSF.Ê The latter
characteristics point towards a past merger with a less massive,
gas-rich galaxy as the source of both the gas and the disturbed
dynamics in the centre of NGC 4150, and as the trigger for the
RSF.Ê

Several authors \citep[e.g.,][]{1982ApJ...257..423H,2001AJ....122.1251K} have associated NGC 4150 with the Canes Venatici I group (also M94 group), the majority of whose member galaxies are within 2-8 Mpc of our Sun, with a median distance of $\sim$4 Mpc \citep{1998A&AS..128..459M}.  However, the distances estimated for NGC 4150 in the literature are somewhat larger than this value.  \citet{2001ApJ...546..681T} calculated a distance modulus, $\mu = 30.69\pm0.24$ (d$\approx13\pm1.5$ Mpc) from measurements of I-band surface brightness fluctuations (SBF), calibrated using Cepheid distances to other nearby galaxies.  \citet{2003ApJ...583..712J} later updated this value to $\mu = 30.53\pm0.24$ after re-calibrating the data using the Cepheid period - luminosity relation of \citet{1999AcA....49..223U}.  \citet{2005MNRAS.361..330R} found $\mu = 30.79\pm0.2$ (d$\approx 14.4$ Mpc) from R-band SBF measurments, while \citet{2003A&A...398..467K} estimated $\mu \approx 31.5$ (d$\sim$20 Mpc) from the turnover magnitude of the globular cluster luminosity function (GCLF).  Here we adopt the mean of the re-calibrated Tonry et al. I-band \citep{2003ApJ...583..712J} and Rekola et al. R-band SBF distance estimates; $\mu = 30.66\pm0.16$ (d = $13.55\pm1.0$ Mpc).  This distance modulus is in good agreement with that derived by \citet{2002MNRAS.329..513D} from the mean heliocentric velocity of the Coma I cloud ($\mu$ = 30.68; assuming $H_{0}$ = 75 km\,s$^{-1}$).  Indeed, de Zeeuw et al. go on to associate NGC 4150 with this group of galaxies, which lies on the outskirts of the Virgo cluster.

We have estimated the stellar mass of NGC 4150, $M_*$, from its $K_s$-band luminosity \citep{2009ApJ...695....1T} using the stellar mass-to-light ratios derived by \citet{2003ApJS..149..289B}.  \citeauthor{2003ApJS..149..289B} derived $M/L$ ratios for a large sample of galaxies from the Two Micron All Sky Survey (2MASS) and the Sloan Digital Sky Survey (SDSS), and compared these with galaxy colours (see \citealt[][Appendix 2]{2003ApJS..149..289B}).  From these $M/L$ - colour relations, and using the $(B - V)$ colour of NGC 4150, we calculated $M_{*}/L_{Ks}$ =  $0.6^{+0.3}_{-0.2}$, and hence $M_{*}$ = $6.3^{+3.1}_{-2.1}\times10^{9}$ $M_{\odot}$.  NGC 4150 is unlikely to harbour an active galactic nucleus (AGN) as just a few ($\sim5$) percent of local galaxies of this mass are classified as AGN \citep{2003MNRAS.341...54K}.

%\citet{2009ApJ...695....1T} measured the star formation rate (SFR) in NGC 4150 to be $\sim4\times10^{-2}$ $M_{\odot}\,{\rm yr^{-1}}$ from Spitzer Space Telescope 24$\,\mu {\rm m}$ observations, and point out that this is greater ($\sim \times4$) than the rate at which mass is being ejected from old stars.  This suggests that stellar mass loss cannot provide sufficient material to sustain the current rate of star formation, and that cold gas must have been accreted from outside the galaxy through, for example, recent mergers.

The mass of molecular hydrogen in NGC 4150 was computed by \citet{2003ApJ...584..260W} and \citet{2007MNRAS.377.1795C} from measurements of CO emission.  The two groups derived masses of $M_{\rm H_2}$ = $3.8\times10^{7}$ $M_{\odot}$ and $6.6\times10^{7}$ $M_{\odot}$ respectively, with much of the discrepancy between these two values being due to different CO-to-H$_2$ conversion factors.

%distance estimates for NGC 4150
% Jensen et al. 2003 I-band Surface Brightness Fluctuations = 12.8 Mpc, mu = 30.53$\pm$0.24
% Tonry et al. 2001 I-band SBF mu = 30.69$\pm$0.24 or d = 13.7 Mpc
% Rekola et al. 2005 R-band SBF mu = 30.79$\pm$0.2 or d = 14.4 Mpc
% Karachentsev et al. 2003 GC luminosity function d $\sim$ 20 Mpc or mu = 31.51
% Tully 1988 mu = 29.93$\pm$0.8 or d = 9.7 Mpc

%molecular gas
%3-4 x 10^7 Msun of H2 (that is molecular hydrogen) gas (Welsh and Sage 2003)
%6.6 x 10^7 Msun of H2 as derived from CO measurements (Combes, Young & Bureau 2007)

In the following sections we introduce new HST WFC3 observations of NGC 4150 and compare its integrated NUV-optical photometry with that of the nearby early-type galaxy population (\S 2), perform a pixel-by-pixel analysis of the central region of the galaxy, estimating the age, mass, metallicity and extinction of the young stellar populations (\S 3), and use our results to determine the star formation and assembly history of this nearby lenticular galaxy (\S 4).

\section{Observations, data reduction and preliminary analysis}
\label{sec:obs_data_prelim_analysis}

Observations of NGC 4150 (see Table~\ref{tab:obs_table}) were made using the newly installed Wide Field Camera 3 (WFC3) on board the Hubble Space Telescope (HST).  The observations formed part of an Early Release Science Program (HST program 11360, PI: Robert O'Connell) led by the WFC3 Scientific Oversight Committee. The main objective of this program is to study star formation in a range of different environments (early-type, quiescent and star-forming galaxies) in the local universe.

All data were downloaded from the HST archive\footnote{http://archive.stsci.edu/hst} at the Space Telescope Science Institute (STScI) via the on-the-fly recalibration (OTFR) pipeline, which implemented the {\em CALWFC3} software to bias, dark and flat-field correct the images.  We further reduced the data locally using the {\em MULTIDRIZZLE} software \citep{2002hstc.conf..337K} to register individual exposures in a given filter, apply distortion corrections, mask out cosmic rays and other defects, and finally combine the exposures using the {\em drizzle} image reconstruction technique developed by \citet{2002PASP..114..144F}.  The data were drizzled using the latest image distortion coefficient tables (IDCTAB) downloaded from the WFC3 reference file website\footnote{http://www.stsci.edu/hst/observatory/cdbs/SIfileInfo/WFC3/reftablequeryindex}.

%\textcolor{blue}{****Perhaps add further details of multidrizzle parameters for UVIS and IR detectors****}

Figure~\ref{fig:BVI_ngc4150} shows a pseudo-colour (RGB) image of NGC 4150 created using the WFC3 UVIS $F814W, F555W$ and $F438W$ frames for the red, green and blue channels respectively.  In this optical image NGC 4150 appears to be a typical S0 galaxy possessing a dominant bulge, albeit with obvious dust lanes towards its centre.  The white boxes indicate the fields-of-view (FOVs) presented in Figure~\ref{fig:n4150_colour_dust}, which shows both pseudo-colour and monochrome images of the central region of NGC 4150.  The RGB images in this case are a combination of UVIS $F657N$ (H$\alpha$+[N II] - red), $F438W$ (green) and $F225W$ (blue), while the greyscale images are {\em unsharp-masked} versions of the $F438W$ data.  Unsharp-masking was performed by firstly median boxcar smoothing the $F438W$ image using a box of 50$\times$50 pixels, before subtracting the smoothed image from the original.  This had the effect of enhancing smaller scale structures, in particular the details of the dust distribution in the centre of NGC 4150.  %\textcolor{red}{***Perhaps should redo this with GALFIT***}
Figure~\ref{fig:n4150_greyscale} shows greyscale versions of the UVIS observations of the galaxy core through each of the $NUV$-optical filters.

Two issues are immediately obvious from the images in Fig.~\ref{fig:n4150_colour_dust} and Fig.~\ref{fig:n4150_greyscale}: (1) the unobscured areas of the core of NGC 4150 are bright in the $NUV$ / $F225W$ (Fig.~\ref{fig:n4150_greyscale}a), appearing blue in the colour composite images (Fig.~\ref{fig:n4150_colour_dust}a\&b); and (2) there are significant amounts of dust, which forms a spiral-like structure and appears to rotate clockwise (North to West) in the plane of the galaxy disk.  Some of this dust lies across the line-of-sight to the very centre and lower part of the galaxy core, blocking our view of this region at shorter wavelengths, as is well illustrated by the sequence of images in Figure~\ref{fig:n4150_greyscale}.

\citet{2009MNRAS.398.2028J} also observed NGC 4150 to have a $NUV$ bright core and {\em blue} $NUV$ - optical colours, albeit at much lower resolution ($NUV$ PSF $\approx6\arcsec$) using the NASA GALEX\footnote{http://www.galex.caltech.edu/} (Galaxy Evolution Explorer) satellite (see Figure~\ref{fig:n4150_GalexHST}).  UV flux is generally associated with two distinct stellar populations; {\em young} ($\lesssim$ 1 Gyr), massive stars, and {\em old} ($\gtrsim$5 Gyr), low-mass, core helium burning (horizontal branch - HB) stars.  The latter are believed to be responsible for the {\em UV-upturn} phenomenon in which evolved early-type galaxies, devoid of young stars, develop a strong UV-excess \citep{1997ApJ...486..201Y}.  UV-upturn is typically observed in cluster elliptical galaxies, a signature of an old population \citep[see][for a review]{1999ARA&A..37..603O}.  Since a UV-upturn requires a galaxy's stellar population to have aged sufficiently to allow development of a strong extreme horizontal branch, we do not expect to observe this phenomenon in galaxies beyond z$\gtrsim$1.5, when the universe was less than 5 Gyr old.  However, NGC 4150 is effectively at z = 0, and hence we must consider that UV-upturn is at least a possibility in this case.

By using the corollary evidence of a central, H$\beta$ absorption ``hotspot'' of roughly 2$\arcsec$ radius \citep{2006MNRAS.369..497K}
- indicative of young ($<$ 2 Gyr) stars -
and the detection of large quantities of molecular gas \citep{2007MNRAS.377.1795C},
- the fuel required to form new stars -
\citet{2009MNRAS.398.2028J} concluded that the $NUV$ flux in NGC 4150 is most likely dominated by {\em young} stars created during a recent period of star formation.

Utilising the exquisite resolution of our HST dataset we can add a further, {\em morphological} argument.  We see in Fig.~\ref{fig:n4150_colour_dust}b and, more clearly, in Fig.~\ref{fig:n4150_greyscale}a that there is significant structure in the $NUV$ image, owing not only to attenuation by dust, but also to the presence of several bright {\em knots} of $NUV$ flux.  %Furthermore, the NUV light is clearly localised to the galaxy core, extending about $3\arcsec$ ($\sim200 pc) from the centre; 1/5 of the effective radius at optical wavelengths.
If the flux was attributable to old, low-mass stars one would expect the $NUV$ image to be much smoother, similar to the optical $V$ ($F555W$) and $I$ ($F814W$) images (disregarding dust attenuation) which effectively trace the old stellar population in NGC 4150.  These localised knots of $NUV$ flux, which exist only within about $3\arcsec$ ($\sim$200 pc) of the centre, point towards a population of {\em younger} stars as their source, the clumps of young stars having not yet dispersed since their formation in the recent past.

%This dust is consistent with the detection of large quantities of molecular hydrogen gas detected in NGC 4150 by \citet{2003ApJ...584..260W} and \citet{2007MNRAS.377.1795C} as discussed in \S\ref{sec:target_n4150} (see Table~\ref{tab:ngc4150_props}).

Figure~\ref{fig:integrated_col_mag} shows a colour-magnitude diagram (CMD) on which we have plotted the integrated UVIS photometry of NGC 4150 within 1 effective radius (R$_{eff}$ $\sim$ 400 pixels) compared with the nearby early-type galaxy (ETG) population identified by \citet{2007ApJS..173..619K} in SDSS DR3 and GALEX MIS data.  The blue horizontal line indicates the $(NUV - V)$ colour of the strongest UV-upturn galaxy in the local Universe, NGC 4552, in which the $NUV$ flux is dominated by HB stars that have evolved from the old, low-mass stellar population. The $(NUV - V)$ colour of NGC 4150 is significantly bluer than this empirical UV-upturn limit, adding further support to the argument that it is due to recent star formation.

%\textcolor{red}{The morphology of the NUV flux (centrally concentrated, knot-like structures) is similar to what we observe in galaxy merger simulations presented by \citep{2009arXiv0912.2629P}.  This adds some evidence supporting a galaxy merger as the trigger for a recent burst of star formation.}

It is worth noting that there is a conspicuous lack of H$\alpha$ emission\footnote{A bright source of H$\alpha$ emission is visible just North of the galaxy core in Fig.~\ref{fig:n4150_greyscale}f.  This is potentially a {\em young globular cluster}, one of a population of such objects in NGC 4150 that will be discussed more fully in a forthcoming paper (Kaviraj et al. in prep).} coincident with the $NUV$ knots, or indeed anywhere in the core of NGC 4150 (Fig.~\ref{fig:n4150_greyscale}e\&f).  The continuum subtracted from the raw $F657N$ image was estimated from an average of the $F555W$ and $F814W$ observations, which was scaled to match the $F657N$ photometry of two K-type stars\footnote{K-type stars were chosen as they exhibit weak H-absorption lines.  Conversely the A-type star visible to the South East of the galaxy core in Figs.~\ref{fig:BVI_ngc4150} and \ref{fig:n4150_colour_dust} was rejected as stars of this type possess strong H-absorption features.} in the field.  The lack of H$\alpha$ emission suggests that there are few, if any, very young stars ($\lesssim$ 10 Myr) in the region and that the proposed burst of recent star formation ended some time ago.  The possibility that NGC 4150 is a {\em post-starburst} system is a point we return to in the following sections.

\section{Detailed analysis - photometry and parameter estimation}

There are several means by which one might attempt to derive the star formation history of a target galaxy or stellar population.  If the constituent stars are sufficiently resolved it may be possible to perform photometry on each in turn, and subsequently derive individual estimates for age, mass and metallicity through comparison with stellar models.  Coeval populations are trivially identified by such analyses and it is possible to build up a detailed and spatially resolved picture of the star formation histories in such cases \citep[e.g.,][]{2008AJ....135.1998R,2000ApJ...533..203S,1994A&A...290...69B}.

Where individual stars are not resolved, one might fit model stellar populations to the integrated photometry of open and/or globular clusters (GCs) within a target galaxy \citep[e.g.,][]{2007MNRAS.381L..74K,2005ApJ...631L.133F}.  In such cases it is generally assumed that the stars within a given cluster formed at about the same time and from the same cloud of gas, and hence they are modelled as simple stellar populations (SSPs) - coeval populations of stars of uniform metallicity.  One can then use the cluster age distribution to make inferences as to the star formation history of the host galaxy.  We perform such an analysis on the globular cluster population in NGC 4150 in a forthcoming paper (Kaviraj et al. in prep.), the results of which are complementary to those presented here.

An alternative method in cases where individual stars are not resolved is to perform photometry on a {\em pixel-by-pixel} basis before fitting stellar population models to each pixel in turn \citep[e.g.][]{2008ApJ...677..970W,2007MNRAS.376.1021J,2003AJ....126.1276K,2003ApJ...586..923E,2003AJ....126.2330C}.  In this way one can build up a contiguous map of the properties of the stellar population within a target galaxy.  It is this method that we employ in the following sections.  While integrated photometry can detect the {\em presence} of different stellar populations, a pixel-by-pixel approach can also define their {\em spatial distribution} within the host galaxy.  The structures revealed in these 2D maps provide vital clues as to where and how the constituent stars formed.

\subsection{Pixel-by-pixel photometry}

Pixel-by-pixel photometry of the core of NGC 4150 was carried out on the WFC3 UVIS observations taken in 5 broadband filters; $F225W, F336W, F438W, F555W \& F814W$.  We have concentrated our efforts on the galaxy core since this is where we observe significant levels of NUV flux, which is most likely associated with young stars.  $NUV$ photometry is crucial for the accurate determination of stellar ages (see \S\ref{sec:param_est}) and hence we limit our analysis to those pixels with $NUV$ ($F225W$) signal-to-noise (S/N) ratios greater than 5.  Small x,y shifts were applied, where necessary, to align each of the drizzled images relative to the $F438W$ frame, before 160$\times$160 pixel square sections of each image were created ($6.4\arcsec$ or 420 pc on each side), centred roughly on the galaxy core.  (The image sections displayed in Fig.~\ref{fig:n4150_greyscale} are in fact those used in the following analysis.)  The sky background in each image was measured from the median counts in several {\em blank} regions of sky to the NE and SW of the galaxy core, and well beyond the extent of its optical disk.  These sky background levels were subtracted from each of the WFC3 images prior to calculating the photometry.%The dimensions of this image section were chosen as it contained all the central pixels with NUV S/N $>$ 5. [x1:x2,y1:y2] = [2125:2284,1910:2069].

The sky-subtracted image sections were read into a custom-built C program which calculated the AB magnitude\footnote{AB$_{\nu} = 0; f_{\nu} = 3.63\times10^{-20}$ ergs s$^{-1}$ cm$^{-2}$ Hz$^{-1}$}, photometric error and S/N ratio of each pixel, applying the zeropoints published on the WFC3 webpages\footnote{http://www.stsci.edu/hst/wfc3/phot\_zp\_lbn}.  Photometric errors were calculated taking into consideration both the total number of photoelectrons collected by a given pixel over the course of an observation, and the readout noise associated with the UVIS detectors - equations~\ref{eqn:flux_err} and~\ref{eqn:phot_err}:

\begin{equation}
\label{eqn:flux_err}
\sigma_{X} = \sqrt{{\rm F}_{X} + {\rm R^2}}
\end{equation}

\begin{equation}
\label{eqn:phot_err}
{\rm m}_{err,X} = -2.5 \,log \frac{{\rm F}_{X} - \sigma_{X}}{{\rm F}_{X}}
\end{equation}

where F$_{X}$ is the total number of photoelectrons collected in a given pixel through bandpass $X$, R is the detector readout noise (UVIS $\sim$3 e$^-$),  $\sigma_{X}$ is the 1 sigma uncertainty in the pixel electron counts as a result of poisson and readout noise, and m$_{err,X}$ is the associated magnitude error.  The signal-ro-noise ratio was also calculated as S/N = F$_{X}$ /  $\sigma_{X}$.

The AB magnitude, photometric error and S/N ratio for each pixel, in each of the five broadband filters, were output to a multi-column text file, while the photometry for each filter was also output in the form of individual FITS images, with S/N thresholds of $\geq$5.

Figures~\ref{fig:pix_phot},~\ref{fig:pix_col_dist} and~\ref{fig:pix_col_col} detail several qualitative analyses of the WFC3 photometry in the core of NGC 4150.  Fig.~\ref{fig:pix_phot} shows a map of the $(NUV - V)$ colour created by subtracting the $F555W$ photometry FITS image from its $F225W$ counterpart.  Pixels with a S/N ratio less than 5 appear white in this map, having been assigned a null value. (In practice the spatial extent of the {\em useful} colour map is limited exclusively by the S/N in the $NUV$ ($F225W$) image.  The S/N ratio of the individual pixels in this image drops below 5 at much shorter radial distance from the galaxy centre compared to the optical bands.)

Figs.~\ref{fig:pix_col_dist} and~\ref{fig:pix_col_col} respectively show a colour distribution $(NUV - V)$ and a colour-colour plot $(NUV - V$ vs $B - V)$ of the core pixels in NGC 4150.  In both cases the NGC 4150 photometry is compared with pixel colours from a star-forming region in M83 (also imaged with WFC3 - \citealt{2010ApJ...719..966C}; plotted in blue), and integrated colours of nearby ETGs \citep[plotted in black;][]{2007ApJS..173..619K}.  Additionally the integrated colours of nearby {\em E+A galaxies} - post-starburst, major-merger remnants with high mass-fractions of recent ($<$ 1 Gyr old) star-formation - are shown on the colour-colour diagram \citep[plotted in green;][]{2007MNRAS.382..960K}.

The central colours of NGC 4150 are consistent with a {\em post-starburst} stellar population, falling between the currently star-forming M83 and the old, passively-evolving ETGs.  This is in good agreement with our previous observation regarding the lack of significant H$\alpha$ emission in the core of NGC 4150 (\S\ref{sec:obs_data_prelim_analysis} and Fig.~\ref{fig:n4150_greyscale}e\&f), which suggests that there are few, if any, very young stars ($\lesssim$ 10 Myr) and that the starburst ended some time ago.

Comparison with the integrated colours of E+A galaxies, {\em `bona-fide'} post-starburst systems, shows that the central pixels of NGC 4150 are much {\em redder} in the $B - V$ (optical) colour, while having similar $NUV - V$ colour.  This can be explained as being due to higher mass fractions of RSF ($\sim$20-60 per cent) in the E+A galaxies \citep{2007MNRAS.382..960K}.  As the mass fraction of RSF increases, the $NUV$-optical colours become dominated by the young stars and therefore change little with further increases in mass fraction (Fig.~\ref{fig:col_vs_mass}).  At the same time, the optical colours become steadily bluer as the optical light from the young stars begins to compete with that from the old, underlying stellar population.  This comparison therefore points towards a lower mass-fraction of recent star-formation in NGC 4150 than in E+A systems, and suggests that any merger event that may have triggered this star-formation was most likely a minor merger (component mass ratio $<$ 1:4).

\subsection{Parameter estimation}
\label{sec:param_est}

We estimated the values of parameters governing the star formation history (SFH) of NGC 4150 by comparing its 5-band pixel photometry $(NUV, U, B, V, I)$ to a library of synthetic photometry,
generated using a large collection of model SFHs, specifically optimized for studying early-type galaxies at low redshift. Our primary aims were to explore the age of the last star formation event at the {\em pixel-by-pixel} level, the fractional mass of stars produced by this star formation, and the metallicity of the recently formed stars.  As a result of the fitting process we also recovered the extinction for each pixel. In this particular case of NGC 4150, the foreground (Galactic) extinction was small (E$(B - V)$ = 0.018; Schlegel et al. 1998).  We therefore chose to not to correct the photometry for foreground extinction prior to fitting, but rather estimate the {\em total} extinction (Galactic + host) using a single extinction law \citep{2000ApJ...533..682C}. This approach might not be reasonable if the foreground extinction was larger, as the extinction curves appropriate for the dust in the Milky Way and host-galaxy could be quite different.  However, the effects of any differences in the extinction curves are mitigated here due to the small value of foreground extinction.

\begin{figure*}
\centering
\includegraphics[width=100mm]{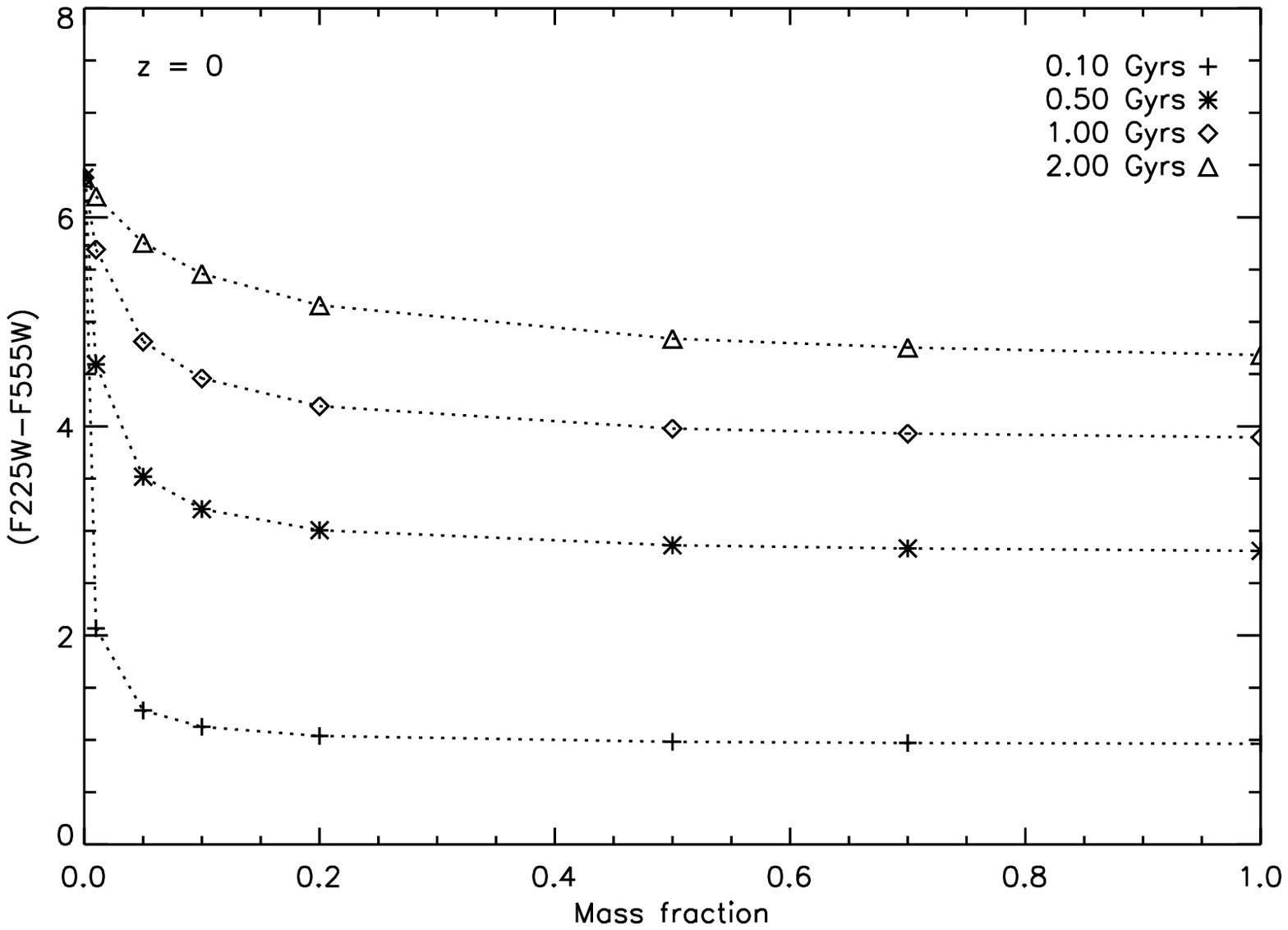}
\caption{Sensitivity of the UV - optical colour of composite stellar populations to young stars.  Each composite population assumes two instantaneous bursts of star formation, the first burst fixed at old age/high redshift (z=3) while the second is free to vary in age and mass fraction.  These models also assume solar metallicity and no dust.  The NUV - optical colour (in this case HST WFC3 $F225W - F555W$) of the composite stellar population is plotted as a function of the age (symbol) and mass fraction (abscissa) of the second burst.  It is clear that even a small mass fraction ($\sim$1 percent) of young stars ($\lesssim$1 Gyr) results in a significant change in the $NUV - V$ colour compared to that of a purely old stellar population ($F225W - F555W \sim 6.4$).  The spectrum of the young component begins to dominate the combined SED as its mass fraction increases beyond $\sim$5 percent.  Hence, for higher mass fractions, the UV-optical colour remains more or less constant.}
\label{fig:col_vs_mass}
\end{figure*}

Before describing our model fitting in detail, we must stress the utility of the UV in alleviating many of the degeneracies that affect similar studies; most notably those between age, metallicity and extinction.  When concerned with the properties of young stellar populations (as we are in this study) the UV spectrum is key, as its response to even small mass fractions of young stars is significantly larger than that of the optical spectrum.  As demonstrated by Fig.~\ref{fig:col_vs_mass} young stellar mass fractions of only a few percent can move the NUV-optical colour of a galaxy bluewards by several magnitudes, away from the red sequence (i.e. a purely old population).  The corresponding change in the optical colours is just a few tenths of a magnitude.  As a consequence, while the optical colours of ETGs are red with very little scatter, their UV-optical colours show a spread of almost 5 magnitudes (see Fig 4 in Kaviraj et al. 2007c). 

Fig.~\ref{fig:col_vs_mass} shows that the size of this bluewards shift in UV-optical colour is extremely sensitive to the age of the RSF, and hence more accurate ages can be determined with the inclusion of UV data.  As demonstrated in Kaviraj et al. (2007b), the addition of UV to traditional optical photometry can effectively break the age-metallicity degeneracy, which allows us to simultaneously estimate both the age and the metallicity of the young stars with a reasonable degree of accuracy and precision.  With regards extinction, the UV-optical ($F225W - F555W$) colour is a factor of four more sensitive to dust than the optical ($F555W - F814W$) colour, which gives us increased leverage in estimating the dust content in the target galaxy. 

We note that, while extremely sensitive to age, the UV-optical colour is only weakly dependent on the mass fraction of the RSF, particularly at mass fractions greater than 5 percent.  This is because at greater mass fractions the young stellar component begins to dominate the galaxy spectrum, in the optical as well as the UV.  The UV-optical colour one measures is then, more-or-less, simply that of the young stars, and increasing the mass fraction further will only slightly affect the measured colour.  However, the availability of the optical spectrum alleviates some of the mass fraction degeneracy since a population with a very large fraction of young stars will have both blue UV and blue optical colours, while a largely old population with a few young stars will have blue UV colours and red optical colours.  The degeneracies between parameters are therefore much less severe with the addition of UV data and this allows us to estimate the age, metallicity and dust content of the target galaxy pixels with a high degree of accuracy.

As we describe below, our scheme decouples the most recent episode of star formation from that which creates the bulk, underlying population. We choose a parametrisation for the model SFHs that both minimises the number of free parameters and captures the macroscopic elements of the star formation history of ETGs in the low-redshift Universe.

Since the underlying stellar mass in ETGs forms at high redshift and over short timescales, we model the bulk stellar population using an instantaneous burst at high redshift. We put this first (primary) instantaneous burst at z=3. Note that small changes to the age of the old population does not affect the derivation of the properties of the young stars. The leverage in the parameters that determine the properties of the recent star formation comes exclusively from the UV spectrum to which the old population does not contribute. The metallicity of the underlying stellar population is fixed at Z = Z$_{\odot}$, since bulk stellar populations in ETGs are observed to be metal rich \citep{1999PASP..111..919H}. Past experience indicates that employing a metallicity distribution \citep[e.g.][]{2007MNRAS.382.1415S} does not alter the derived values of RSF parameters compared to employing an old stellar population with a single metallicity \citep[e.g.][]{2007ApJS..173..619K}.

A large body of recent evidence suggests that the star formation in these systems in the \emph{local} Universe is driven by minor mergers \citep[see][and references therein]{2010arXiv1001.2141K,2010arXiv1004.3775K}. This star formation is bursty and we model the RSF episode using a second instantaneous burst, which is allowed to vary in age between 0.001 Gyrs and the look-back time corresponding to $z=3$, and in mass fraction between 0 and 1. Our parametrisation is similar to previous ones used to study elliptical galaxies at low redshifts \citep[e.g.][]{2000ApJ...541L..37F}.

To build the library of synthetic photometry, the metallicity of the second instantaneous burst of star formation is allowed to vary in the range 0.04Z$_{\odot}$ to 2.5Z$_{\odot}$, while a value of dust extinction, parametrised by E$(B - V)$ in the range 0 to 1.0, is added to the combined model SFH.  The dust model employed in this study is the empirical dust prescription of \citet{2000ApJ...533..682C}. Photometric predictions are generated by combining each model SFH with the chosen metallicity and E$(B - V)$ values and convolving with the stellar models of \citet{2003ApJS..144..259Y} through the WFC3 filtersets.  The uncertainties in the stellar models \citep[which may contribute to offsets from observational data e.g.][]{2001AJ....122.2267E,2003ApJ...582..202Y,2009A&A...493..425M} are taken to be 0.05 mags for the optical filters and 0.1 mags for the $NUV$ passband. The model library contains $\sim750,000$ individual models.

The primary free parameters in this analysis are the age ($t_2$), metallicity ($Z_2$) and mass fraction ($f_2$) of the second burst (the mass fraction of the primary burst is simply $1-f_2$). A secondary parameter of interest is the overall dust properties of the system.  In each case, the value of the free parameters are estimated by comparing the photometry of each pixel to every model in the synthetic library, with the likelihood of each model ($\exp -\chi^2/2$) calculated using the value of $\chi^2$, computed in the standard way. From the joint probability distribution, each parameter is marginalised to extract its one-dimensional probability density function (PDF). We take the median of this PDF as the best estimate of the parameter in question and the 16 and 84 percentile values as the `one-sigma' uncertainties on this
estimate. In the analysis that follows we present these median parameter values and one-sigma uncertainties.

Note that the quality of the $t_2$ fits depends critically on our access to the rest-frame UV, which hosts most of the flux from hot, young main sequence stars. The leverage in $t_2$ comes entirely from the UV/optical colours.  Adding more long-wavelength (e.g. WFC3 IR) filters has no impact on the estimation of RSF
parameters.  Prior experience with 2MASS and UKIDSS NIR data indicates that addition of NIR filters makes the PDFs {\em broader}, but does not change the median values of the parameters themselves.  \citet{2009ApJ...701.1839M} also found that inclusion of NIR data had little or no effect on mean parameter values.

\section{Results and Discussion}
\label{sec:results}

Figures~\ref{fig:param_results}\&\ref{fig:mass_fraction} shows the results from our parameter fitting procedure in the form of both histogram plots of the one-dimensional distributions of the median parameter values within pixels, and as two-dimensional parameter maps.  In each case the plots represent the parameters of the second and {\em most recent} burst of star formation in a given pixel, while the underlying population is assumed to have formed at high redshift (z = 3) and be of solar metallicity (see \S\ref{sec:param_est}). Below we discuss the results for each of the parameters (age, metallicity, extinction and mass-fraction) in turn, and offer our interpretation in the context of the star formation and merger history of NGC 4150.

\subsection{Age}
\label{sec:age}

Figures~\ref{fig:param_results}a\&b respectively show the 1D distribution and the 2D map of the age of the most recent burst of star-formation in the central pixels of NGC 4150.  The age histogram reveals a narrow peak in the RSF age around 0.9 Gyr\footnote{To check the consistency of our technique, in Fig.~\ref{fig:param_results} we compare the NGC 4150 pixel age distribution with that of a currently star forming region in M83. As one would expect we see a clear peak close to zero for the M83 distribution.}(1$\sigma$ uncertainty of $\pm$0.12 Gyr), confirming NGC 4150 to be a post-starburst system as was qualitatively suggested by the lack of H$\alpha$ emission (Fig.~\ref{fig:n4150_greyscale}) and the similarity of $NUV$-optical colours of the central pixels to those of E+A galaxies (Fig.~\ref{fig:pix_phot}).  We further interpret this age as that of an assumed merger event which we believe both supplied raw material to, and triggered, the recent starburst.  %The narrow width of the peak in the pixel age distribution suggests an efficient burst of star formation, truncated within a few 100 Myr such that it appears quasi-instantaneous.  However, we note that we have fitted the RSF in each pixel as an instantaneous burst, and hence this narrow peak is more accurately interpreted as 

The RSF age map reveals further details.  The vast majority of pixels, as we know from the 1D histogram plot, have an age of 0.8 - 1.0 Gyr, which are plotted as light blue / turquoise on the 2D map.  This Gyr old population forms a broad, spiral-like structure which appears to sweep clockwise from the SE (bottom left) around to the north and finally into the core of the galaxy.  It is possible that this population actually forms a ring around the galaxy core, the southern part of which is hidden by dust.  However, superimposed on the broad Gyr old population is a narrow, clumpy stream of young stars ranging from $\sim$500 Myr (dark blue) to $\sim$50 Myr (purple), which spirals from the NE (top left) clockwise into the galaxy core.  The age of the RSF varies along this stream, becoming younger as it approaches the core, while the entire feature appears to rotate in the same direction as the dust visible in the unsharp-masked $F438W$-band image.  %\textcolor{red}{This bears some resemblance to a tidal tail left behind by a minor merger.}
The clumps ($\sim$5 - 10 pc diameter) appear to be large clusters of young stars.  It is possible that the 1 Gyr old population formed in a similar way , the clusters having since dissolved to form the more diffuse structure we see today.

%Many of the knots of NUV flux visible in Fig.~\ref{fig:n4150_greyscale}a can be identified with the young (50 to 500 Myr), clumpy substructure in the age map.  However, by first correcting the NUV image using the extinction map in Fig.~\ref{fig:param_results}f, we can compare the intrinsic NUV light to the age map and find a much closer correlation.

\subsection{Metallicity}
The median pixel values and spatial distribution of the metallicity of the young stars are shown in Figures~\ref{fig:param_results}c\&d.  The metallicity distribution shows a peak at sub-solar metallicity, with a median value of $\sim$0.5 Z$_{\odot}$ and is truncated at 0.3 Z$_{\odot}$ (1$\sigma$ uncertainty of $\pm$0.13Z$_{\odot}$).  Note that our model library included metallicities as low as 0.04 Z$_{\odot}$.

From the 2D metallicity map there is a clear evidence for an increase in metallicity towards the centre of the galaxy, with values in the central region (yellow to red) ranging from roughly 1.0 to 1.7 Z$_{\odot}$.  Along the clumpy stream of young stars mentioned in \S~\ref{sec:age}, we see an increase in metallicity with decreasing age, with some of the highest metallicities measured being associated with the very youngest populations of stars ($\sim$30 - 100 Myrs - purple in age map).  However, some pixels on the southern edge of the very young (purple) feature near the centre of the age map have metallicities of between 0.3 - 0.7 Z$_{\odot}$.  Indeed the youngest `pixel', with an RSF age of just 4 Myr, has a metallicity of 0.29 Z$_{\odot}$.

The Gyr old population (light blue / turquoise in age map) which forms the broad, perhaps spiral, structure we described in the previous section is of sub-solar metallicity, in the range of 0.3 - 0.7 Z$_{\odot}$.  However, there appears to be a {\em distinct} population of 0.8 - 1.0 Gyr stars, either side of the dust-obscured core, with greater than solar metallicity (orange/red pixels either side of core in metallicity map, which appear light blue/turquoise in age map).  Note that the difference in metallicity between these populations (0.3 - 0.7 Z$_{\odot}$ and 1.2 - 1.7 Z$_{\odot}$) is significant when compared to the typical 1$\sigma$ uncertainty of $\pm$0.13Z$_{\odot}$.  In Fig.~\ref{fig:metal_pdf} we present, separately, marginalized PDFs for pixels which have young-stars with (median) metallicities less than 0.4 Z$_{\odot}$ and young-stars with (median) metallicities between 1 and 1.5 Z$_{\odot}$ respectively.  It is clear from these plots that the low and high metallicity populations are indeed distinct.  This clear separation is made possible due to the smaller degeneracies between parameters which result from the inclusion of the UV data (see \S\,\ref{sec:param_est}).  These two populations clearly formed at around the same time, but must have formed from two distinct sources of gas with different metallicities.  We propose that the broad, low-metallicity population formed from material accreted from a gas-rich satellite galaxy during a recent merger/interaction, while the higher-metallicity, central population may have formed from metal-rich gas already present in NGC 4150, which, due to gravitational torques induced by the merger/interaction, was caused to lose angular momentum and fall to the galaxy centre.

Assuming that our hypothesis is correct, the truncated peak in the metallicity distribution (Fig.~\ref{fig:param_results}c) is then representative of the gas-phase metallicity of the galaxy that merged with NGC 4150 roughly 1 Gyr ago.  From the mass-metallicity relation of \citet{2004ApJ...613..898T} we estimate the mass of the accreted galaxy to be $\sim$3$\times10^8$ M$_{\odot}$, roughly 1/20 the mass of NGC 4150, and consistent with a minor merger.  This is admittedly a very crude estimate.  The extrapolated 1$\sigma$ uncertainties from \citealt[Fig.6 of ]{2004ApJ...613..898T} yield masses for the accreted galaxy in the range of 1 - 8$\times10^8$ M$_{\odot}$, assuming Z = 0.5Z$_{\odot}$.  Combining these values with the lower limit for the stellar mass of NGC 4150 from Table~\ref{tab:ngc4150_props} (4.2$\times10^9$ M$_{\odot}$) we find an {\em upper limit} for the merger component mass ratio of $\sim$ 1:4.25, which is close to the threshold (1:4) but still consistent with a minor merger.

Previous observational studies and N-body simulations have found that mergers and interactions can lead to flattening of the metallicity gradients in the component galaxies, mainly due to inflows of metal-poor gas from the outskirts of each galaxy \citep[e.g.][]{2010ApJ...710L.156R, 2009A&A...499..427D, 2008MNRAS.389.1593K, 2008MNRAS.386L..82M, 2000ApJ...538..141M}.  Unfortunately, we can infer little about the global metallicity gradient in NGC 4150 from our HST analysis, as the UV flux is restricted to just the core of the galaxy.  However, we recognize that these studies suggest an alternative source for some (if not all) of the low-metallicity gas which fueled most of the RSF in the galaxy core; i.e. the inflow of metal-poor gas from the outer parts of NGC 4150 during the galaxy merger/interacton.  This, of course, requires that NGC 4150 harboured such a gas reservoir in the first place.  We also note that if this scenario is correct, and the gas fueling the RSF was sourced mostly (or entirely) from within NGC 4150, we cannot use the metallicity of the metal-poor RSF to constrain the mass of the companion galaxy.

%This is admittedly a very crude estimate (the extrapolated 1$\sigma$ uncertainties from \citealt[][Fig. 6]{2004ApJ...613..898T} yield masses in the range of 1 - 8$\times10^8$ M$_{\odot}$, assuming Z = 0.5Z$_{\odot}$) but it is the best that can be achieved with the available data.

\subsection{Extinction}
Figures~\ref{fig:param_results}e\&f show the distribution and map of extinction values for each of our fitted pixels.  Note that the pixel-by-pixel photometry was not corrected for small Galactic extinction  (E$(B - V)$ = 0.018; Schlegel et al. 1998) prior to running the parameter estimation (see \S\,\ref{sec:param_est}).  The typical 1$\sigma$ uncertainty on the fitted values of E$(B - V)$ is $\pm$0.035.

The most obvious (and again reassuring) observation from the extinction map is that it tends to trace the dust, with higher levels of extinction being associated with the dustier regions observed in Figs.~\ref{fig:n4150_colour_dust}\&\ref{fig:n4150_greyscale}.  The highest levels of extinction (E$(B - V) \sim$ 0.6 - 0.8) are associated with the most recently formed stars ($\sim$30 - 100 Myrs), which are found in the very core of the galaxy (purple in age map).  However, we note that the central population of Gyr old stars, with super-solar metallicities, suffers significantly lower levels of extinction (E$(B - V) \sim$ 0.1 - 0.3).

\subsection{Mass Fraction}

%\textcolor{red}{The mass fraction distribution and map (Figures~\ref{fig:mass_fraction}) are a little less clear.  However, there seems to be some evidence that the young ($\lesssim$ 500 Myr), clumpy features in the age map have relatively small mass fractions, ranging from roughly 0.1 - 10 percent per pixel.  The smoother features from the age map have mass fractions of 20 to 80 percent.  ***More text here***  Approximately 12 percent of V-band galaxy light is contained within the fitted pixels in the core of NGC 4150.  Assuming a constant mass-to-light ratio for the entire galaxy, this further suggests that roughly 12 percent of the galaxy mass is contained within these pixels.  The mean mass fraction of RSF per pixel is roughly 0.35.  We therefore estimate that the mass of recently formed stars constitutes roughly 4 percent of the mass of NGC 4150.}

Figure~\ref{fig:mass_fraction} shows the mass distribution and map. Note that the mass fraction uncertainties are typically larger because the same UV colour may be consistent with a wide range of mass fractions. This is essentially because as the mass fraction increases the young stellar component begins to dominate the SED, so that the normalisation changes but the shape of the SED (which determines the colours) does not. While the UV colour changes rapidly with age (regardless of the mass fraction) the mass fraction itself is more degenerate (Fig.~\ref{fig:col_vs_mass}). Typically the mass fraction errors are much better constrained for low values of mass fractions, which correspond to younger ages ($f_{2,err}$ = $\pm$0.04 for $f_2 < 0.2$; $f_{2,err}$ = $\pm$0.1 for $f_2 > 0.2$).

Weighting the pixel mass fractions by their uncertainties yields a typical of RSF mass fraction of around 20 percent per pixel.  We find that approximately 12 percent of V-band galaxy light is contained within the fitted pixels in the core of NGC 4150, which further suggests (assuming a constant mass-to-light ratio for the entire galaxy) that roughly 12 percent of the galaxy mass is contained within these pixels.  We therefore estimate the {\em total} mass fraction of young stars in this galaxy to be around 2-3 percent.

Note that the typical values for the age (1 Gyr) and mass fraction (20 percent) of young stars in the central pixels of NGC 4150 implies that the young stellar component contributes around 94 percent of the flux in $F225W$ ($NUV$), and 64 percent of the flux in $F555W$ ($V$) across the fitted region.

%One might be able to estimate a lower limit to the mass of the accreted galaxy by combining the mass of the RSF with the molecular hydrogen mass. This does not take into account the mass of the accreted galaxy that was in old stars, which simulations show will be distributed is shells throughout the merged galaxy.

%**Other points to discuss***
%\textcolor{blue}{I am uncertain as to how to properly interpret the clumpy stream of 500 - 50 Myr stars discussed earlier.  I am assuming it is caused by stars forming sequentially from gas as it spirals into the centre of the galaxy.  As I have already said, the age of the most recent SF in these pixels decreases along the spiral feature towards the core, while metallicity increases.  Is it possible that the gas is being enriched by successive generations of massive stars as it spirals into the centre of the galaxy, forming stars en route from ever more metal rich material?}

\section{Conclusions}

\begin{figure*}
\centering
\includegraphics[width=80mm]{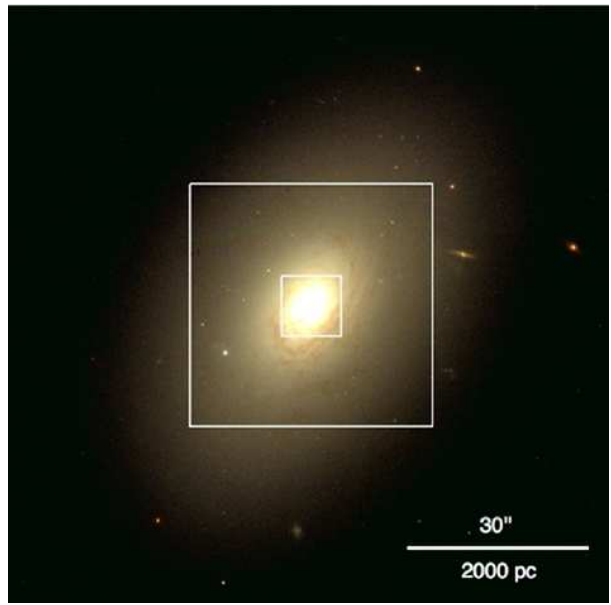}
\caption{RGB image of local S0 galaxy NGC 4150 created from WFC3 UVIS $F438W$ (Blue), $F555W$ (Green) and $F814W$ (Red) observations.  The white squares indicate the fields-of-view shown in Fig. 2a\&c (large box) and Fig. 2b\&d (small box).  North is up and East is to the left.}
\label{fig:BVI_ngc4150}
\end{figure*}

\begin{figure*}
\centering
\begin{tabular}{cc}
\includegraphics[width=60mm]{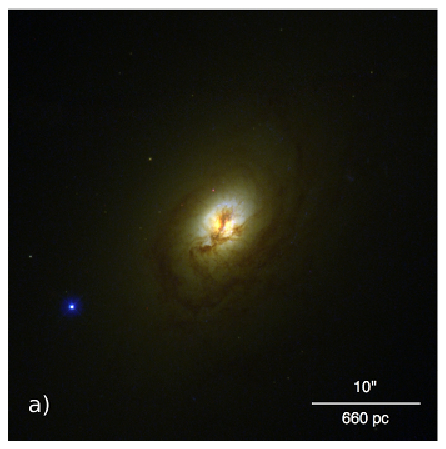}&
\includegraphics[width=60mm]{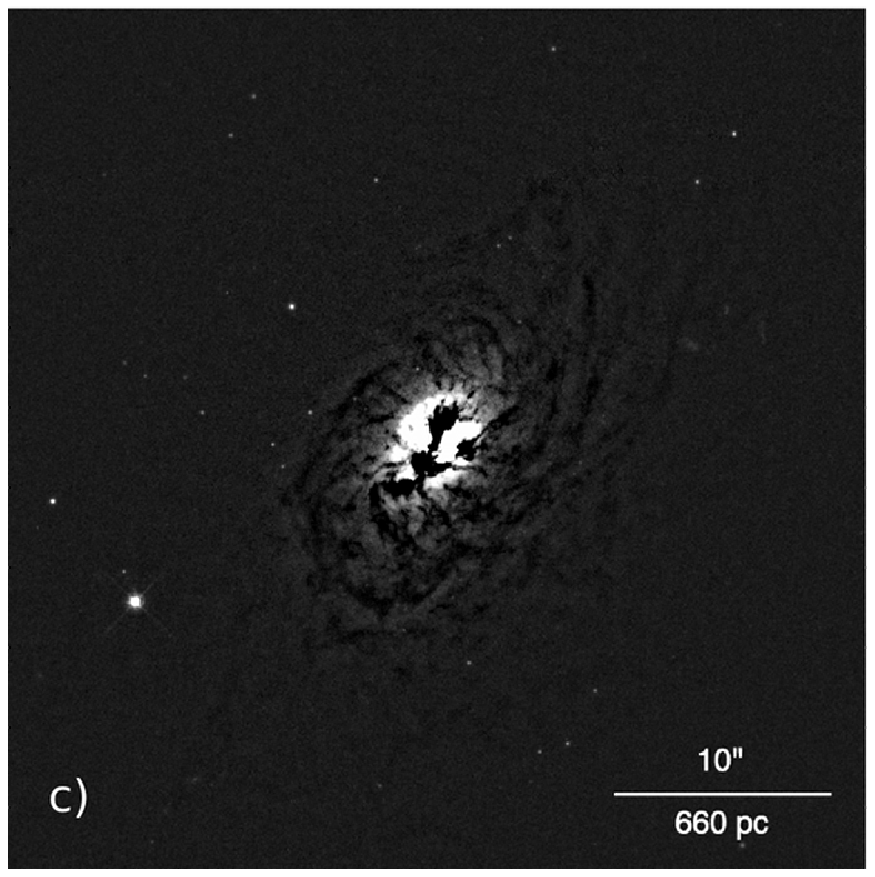}\\
\includegraphics[width=60mm]{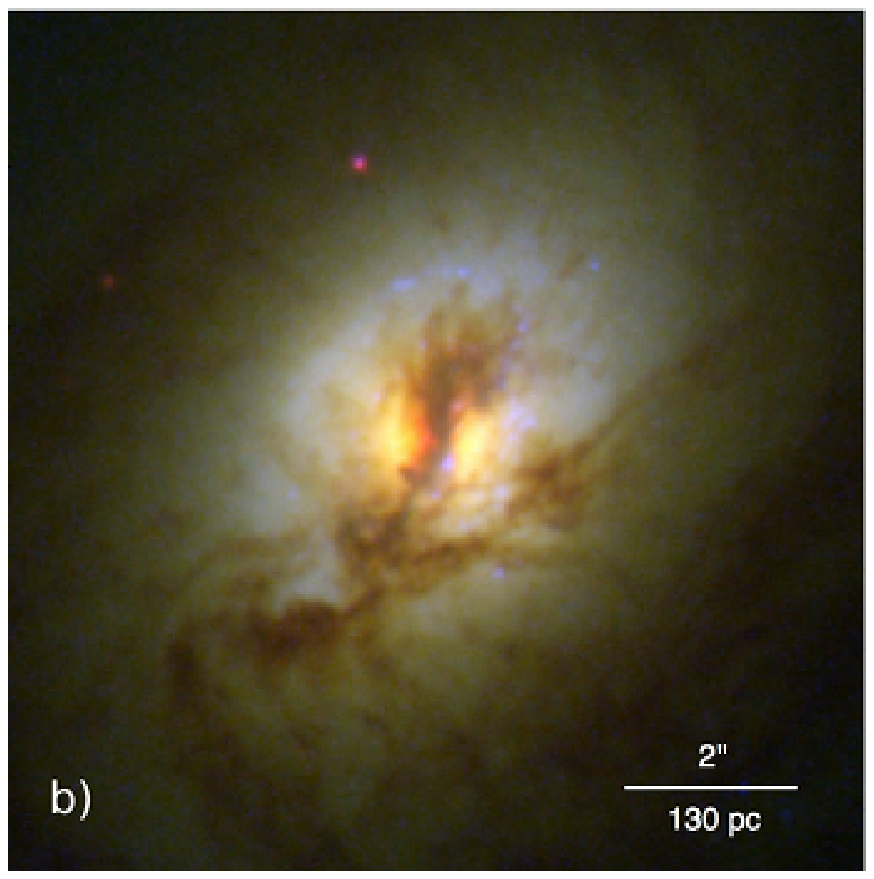}&
\includegraphics[width=60mm]{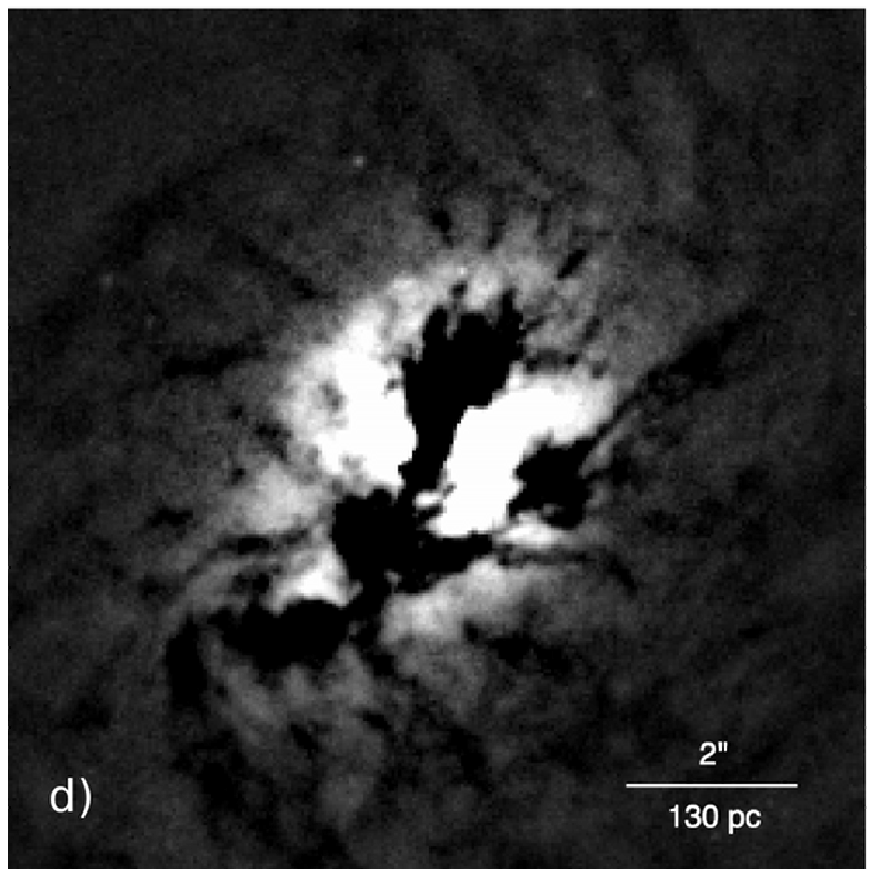}
\end{tabular}
\caption{HST WFC3 images of NGC 4150 (fields-of-view indicated on Fig.1).  {\bf (a\&b)}: RGB images created using $F225W$ (Blue), $F438W$ (Green) and $F657N$ (Red) WFC3 UVIS observations.  The areas of the core not obscured by dust are bright in $F225W$ (NUV) indicative of a relatively young population of stars.  {\bf (c\&d)}: Unsharp-masked $F438W$ images revealing the distribution of dust.  In all cases North is up and East is to the left.}
\label{fig:n4150_colour_dust}
\end{figure*}

\begin{figure*}
\centering
\begin{tabular}{cccc}
\includegraphics[width=35mm]{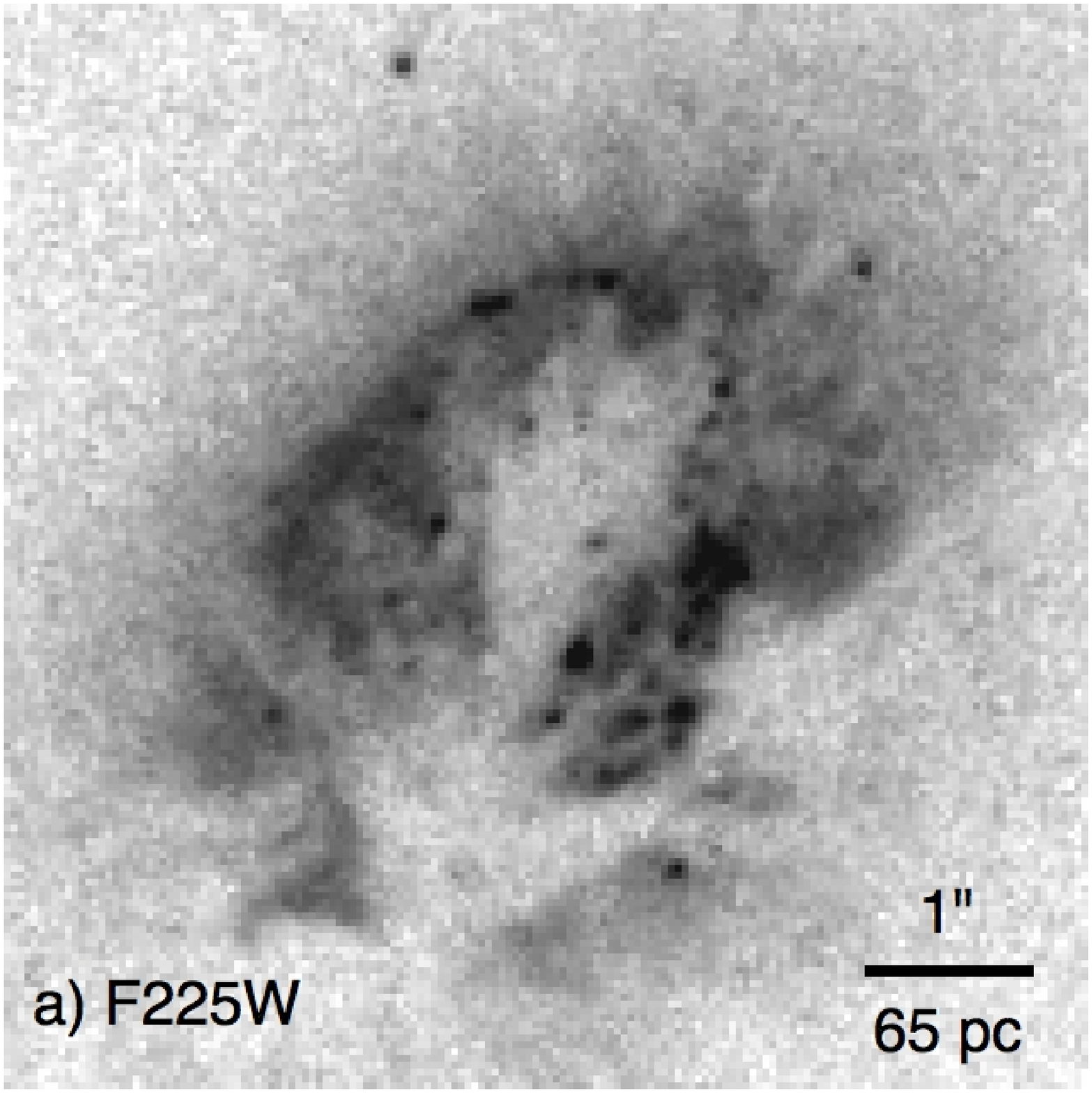}&
\includegraphics[width=35mm]{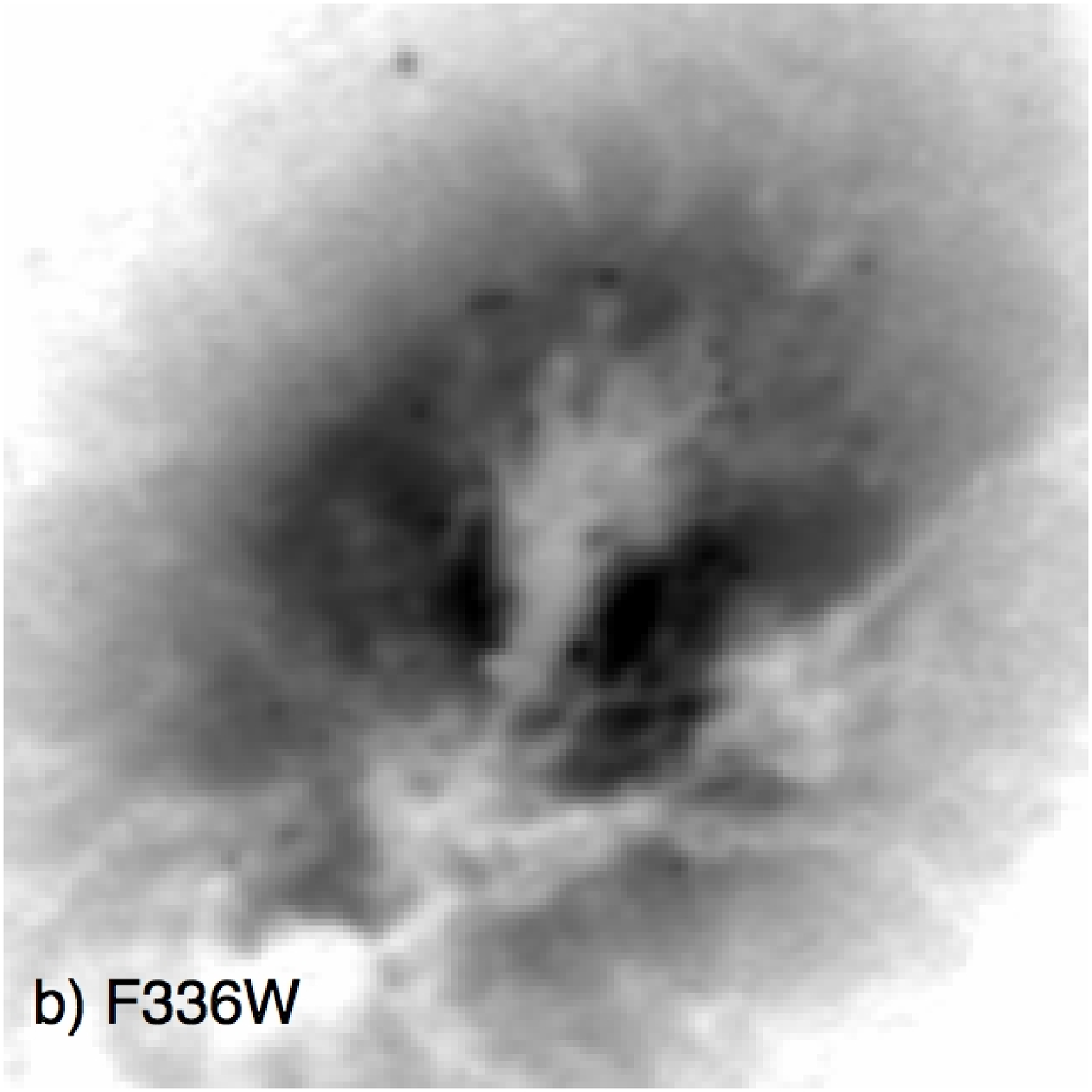}&
\includegraphics[width=35mm]{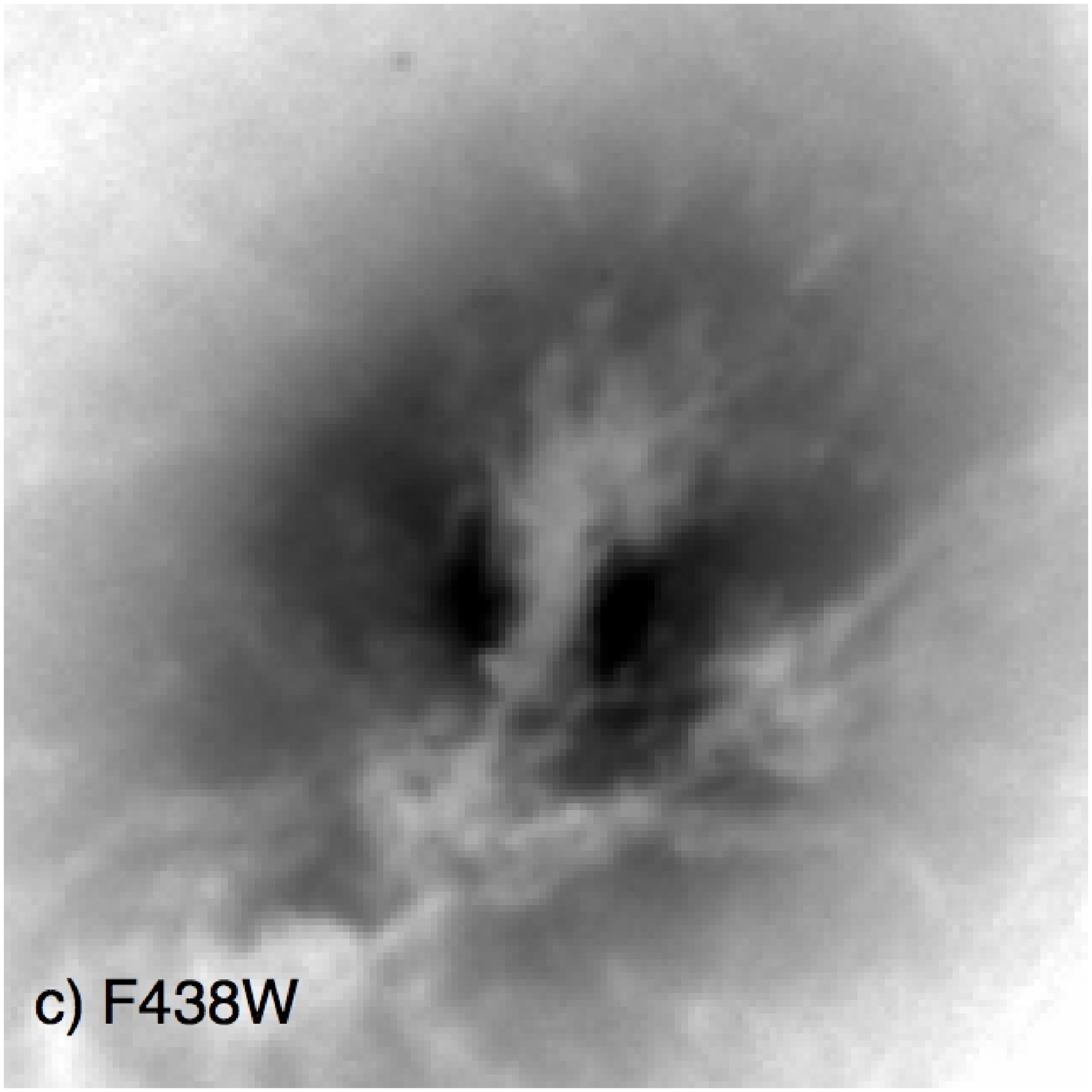}&
\includegraphics[width=35mm]{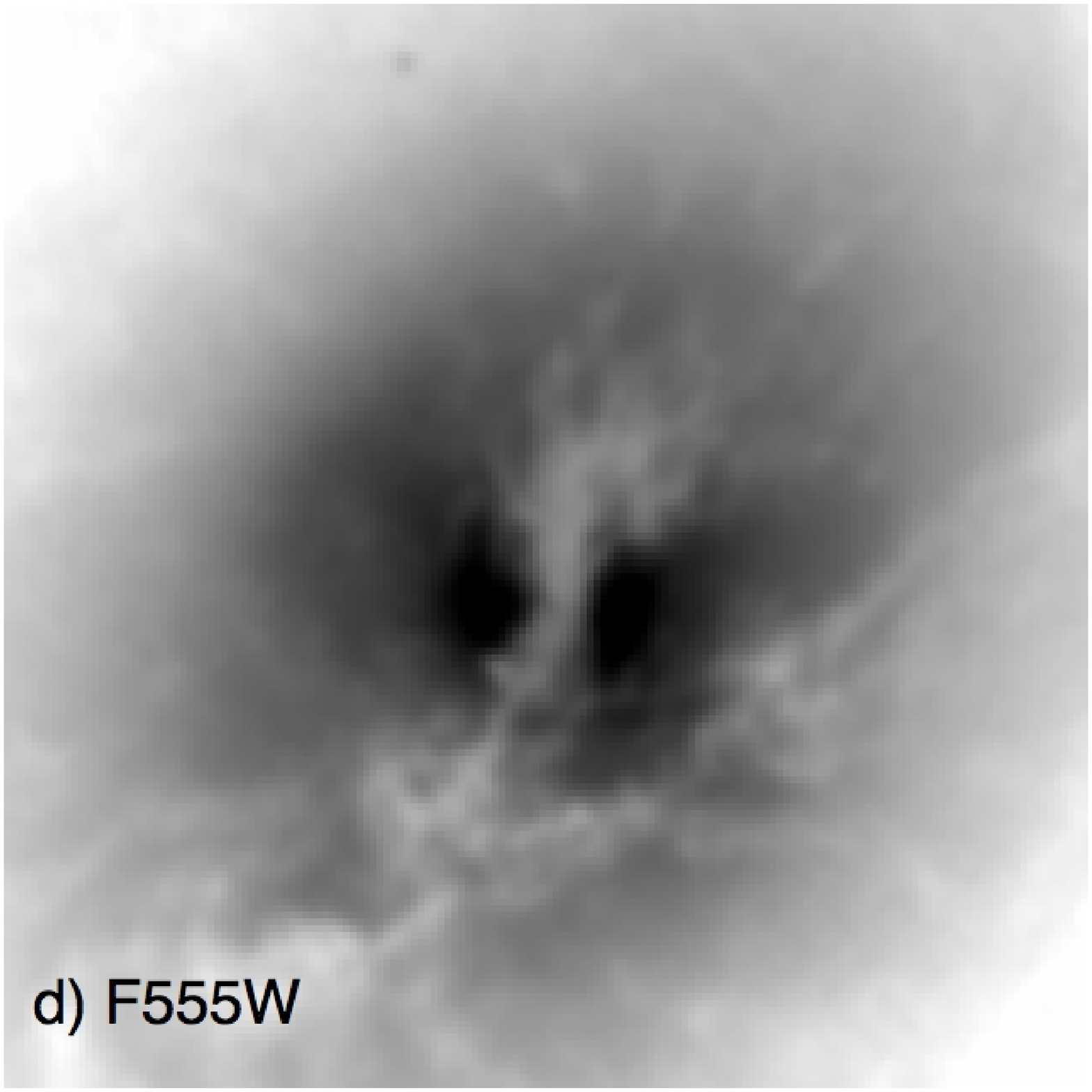}\\
\includegraphics[width=35mm]{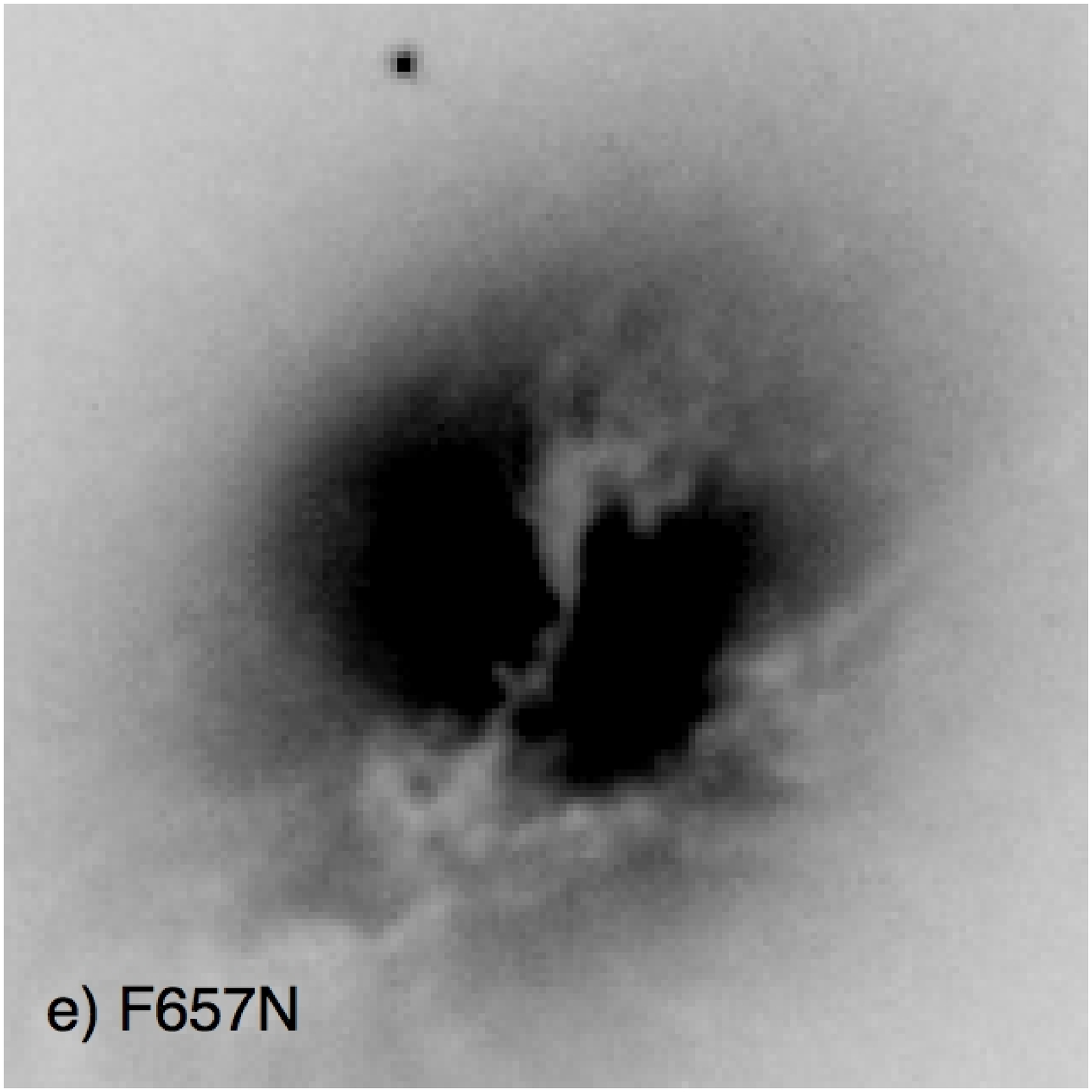}&
\includegraphics[width=35mm]{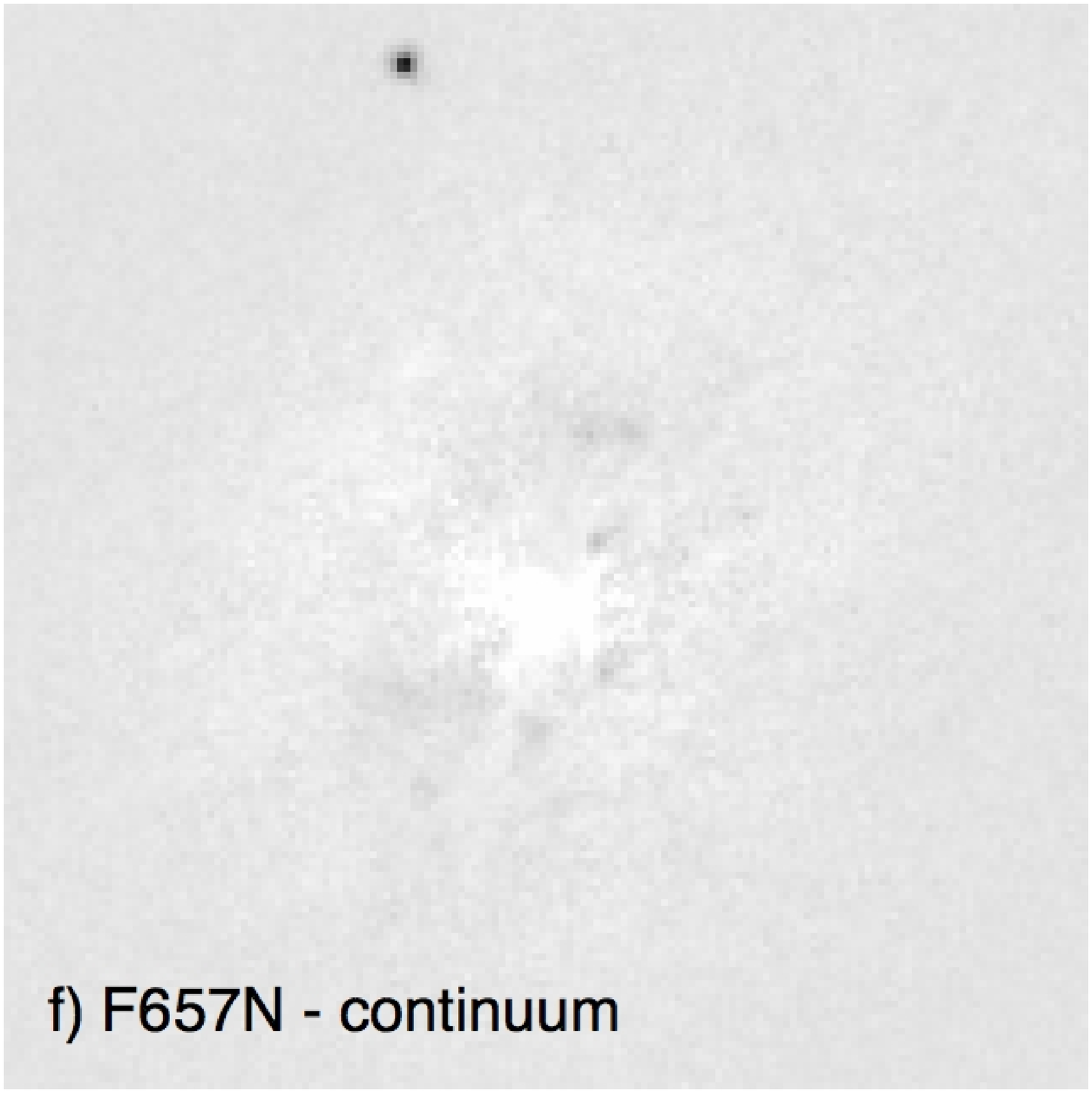}&
\includegraphics[width=35mm]{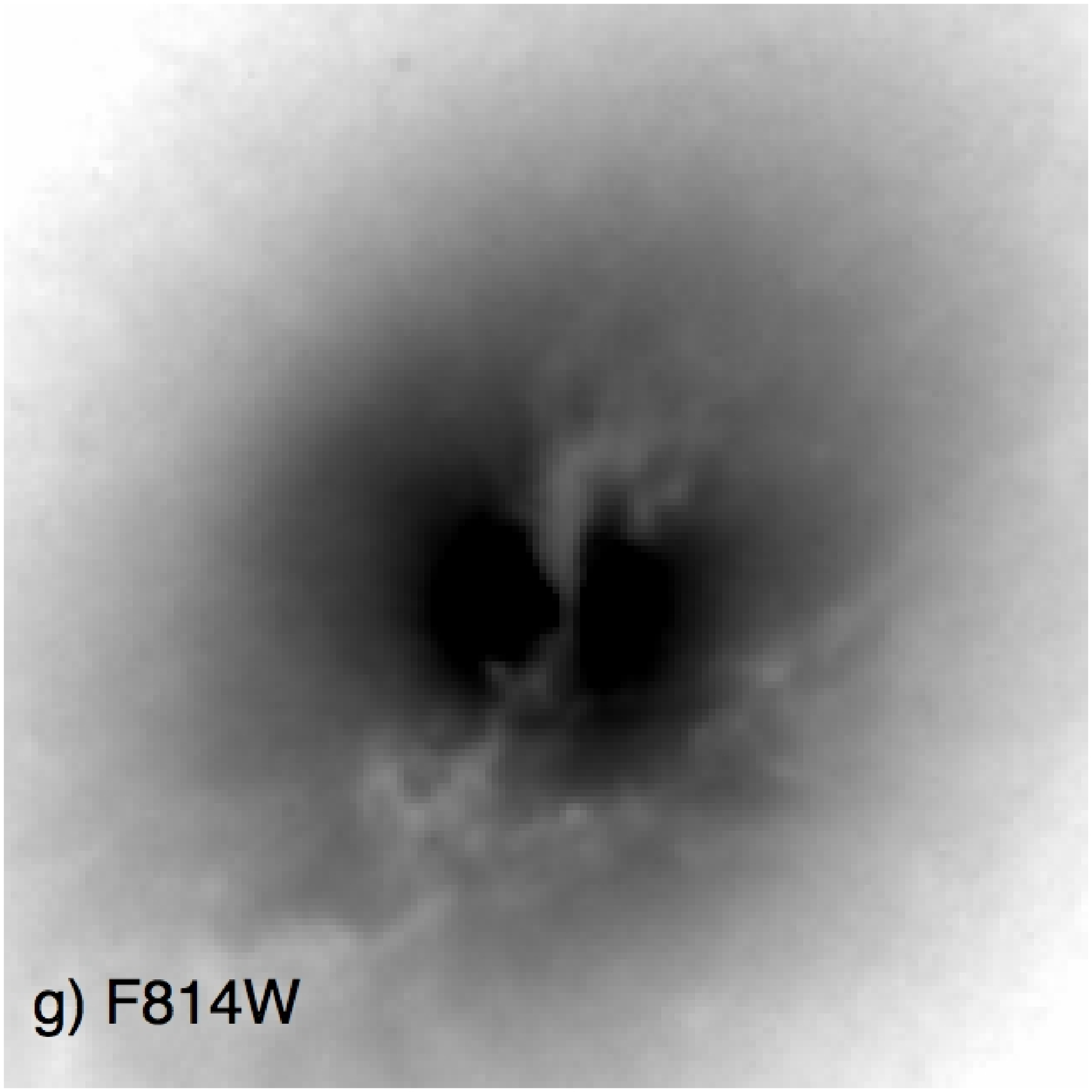}&
\end{tabular}
\caption{Individual bandpass HST WFC3/UVIS images of the core of NGC 4150.  {\bf (a-f)}: $F225W, F336W, F438W, F555W, F657N$ (Wide H$_{\alpha}$+ [N II]), $F657N$ {\em minus} continuum (which has the same flux scaling as $F657N$), and $F814W$.  Note the clear presence of structure in the $NUV$ ($F225W$) image, indicative of recent star formation.  Note also the lack of significant H$_{\alpha}$ emission in the $F657N$ - continuum frame, suggesting that there are few very young stars of $<$ 5 - 10 Myrs.  All images oriented such that North is up and East is to the left.}
\label{fig:n4150_greyscale}
\end{figure*}

\begin{figure*}
\centering
\includegraphics[width=120mm]{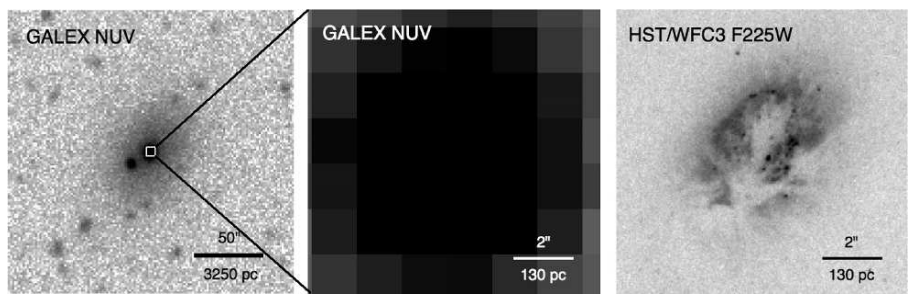}
\caption{Comparison of GALEX and HST/WFC3 observations of NGC 4150, both taken in the $NUV$.  The GALEX $NUV$ PSF (6$\arcsec$) is $\sim$20,000 times larger in area than that of the properly drizzled WFC3 images (0.04$\arcsec$).  As a result the entire central region of NGC 4150, which we see in exquisite detail in the WFC3 data, is completely unresolved by GALEX.}
\label{fig:n4150_GalexHST}
\end{figure*}

\begin{figure*}
\centering
\includegraphics[width=80mm]{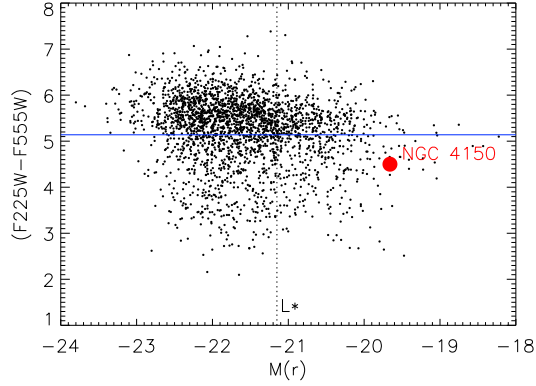}
\caption{Colour-magnitude diagram showing integrated photometry of NGC 4150 within 1 R$_{eff}$ compared with the nearby early-type galaxy (ETG) population identified by \citet{2007ApJS..173..619K} in SDSS DR3 and GALEX MIS data.  The blue horizontal line indicates ($NUV - V$) colour of the strongest UV-upturn galaxy in the local Universe, NGC 4552, in which the $NUV$ flux is dominated by the old stellar population (extreme horizontal branch stars). The ($NUV - V$) colour of NGC 4150 is significantly bluer and is most likely due to recent star formation.}
\label{fig:integrated_col_mag}
\end{figure*}

\begin{figure*}
\centering
\includegraphics[width=60mm]{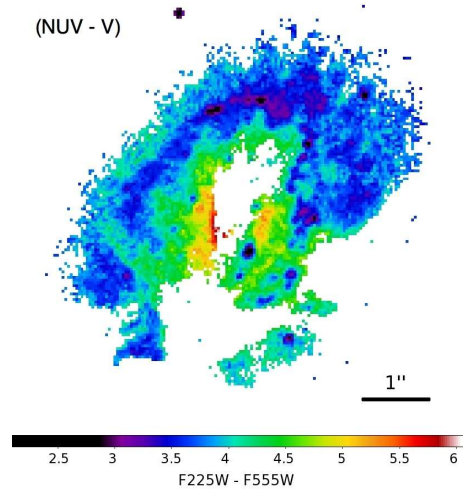}
\caption{NUV - V colour map of central 6$\arcsec$ x 6$\arcsec$ of NGC 4150 using a signal-to-noise cut of 5$\sigma$.  The colour distribution and colour-colour plots of these pixels are shown in Figures~\ref{fig:pix_col_dist}\,\&\,\ref{fig:pix_col_col}.}
\label{fig:pix_phot}
\end{figure*}

\begin{figure*}
\centering
\includegraphics[width=80mm]{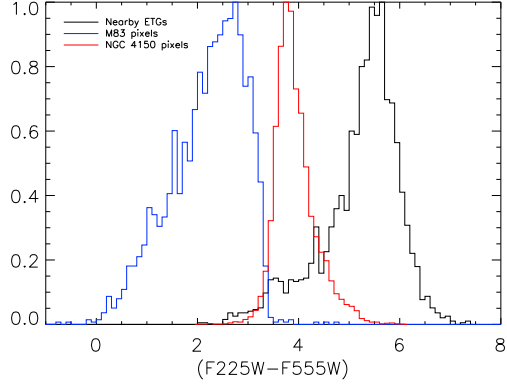}
\caption{Colour distribution of pixels in the central region of NGC 4150 (red) compared with pixel colours from a star-forming region in M83 (blue - also imaged with WFC3) and integrated colours of nearby ETGs \citep[black;][]{2007ApJS..173..619K}.  The central colours of NGC 4150 are consistent with a post-starburst stellar population, falling between the currently star-forming M83 and the old, passively-evolving ETGs.  See also Fig.~\ref{fig:pix_col_col}}
\label{fig:pix_col_dist}
\end{figure*}

\begin{figure*}
\centering
\includegraphics[width=80mm]{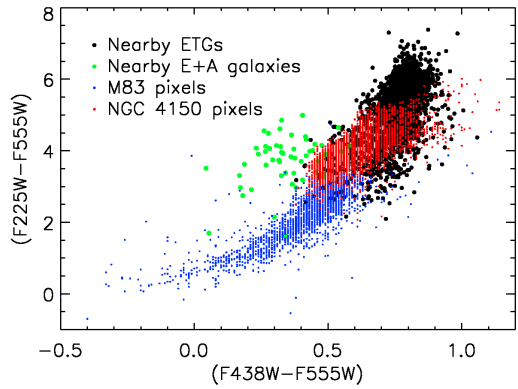}
\caption{Colour-colour plots of pixels in the central region of NGC 4150 (red) compared with pixel colours from a star-forming region in M83 (blue - also imaged with WFC3), integrated colours of nearby ETGs \citep[black;][]{2007ApJS..173..619K}, and integrated colours of nearby E+A galaxies \citep[green;][]{2007MNRAS.382..960K} - post-starburst, major-merger remnants, with high mass-fractions of recent ($<$ 1 Gyr) star-formation.  The central colours of NGC 4150 are consistent with a post-starburst stellar population, falling between the currently star-forming M83 and the old, passively-evolving ETGs in colour-colour space.  Redder optical colours (F438W-F555W) than E+A galaxies suggest a lower mass-fraction of recent star-formation in NGC 4150.}
\label{fig:pix_col_col}
\end{figure*}

We have presented a spatially-resolved NUV-optical study of the early-type (S0) galaxy NGC 4150, using new HST WFC3 data.  Previous work by other authors has shown this galaxy to have a stellar mass of around $6\times10^9$ M$_{\odot}$, roughly $5\times10^7$ M$_{\odot}$ of molecular hydrogen, a kinematically decoupled core (indicative of recent merging) and strong, central H$\beta$ absorption (indicative of young stars).  While relatively uninspiring in its optical image (Fig.~\ref{fig:BVI_ngc4150}), the core of NGC 4150 shows ubiquitous $NUV$ emission and remarkable dusty substructure (Figs.~\ref{fig:n4150_colour_dust}\,\&\,\ref{fig:n4150_greyscale}). The galaxy lies in the UV {\em green valley} (Fig.~\ref{fig:integrated_col_mag}) and its pixels exhibit a narrow range of $(NUV - V)$ colours that are similar to those of nearby {\em post-starburst} (E+A) galaxies, and lie between those of M83 (an actively star-forming spiral) and the local quiescent ETG population (Fig.~\ref{fig:pix_phot}).

\begin{figure*}
\centering
\begin{tabular}{cc}
\includegraphics[width=70mm]{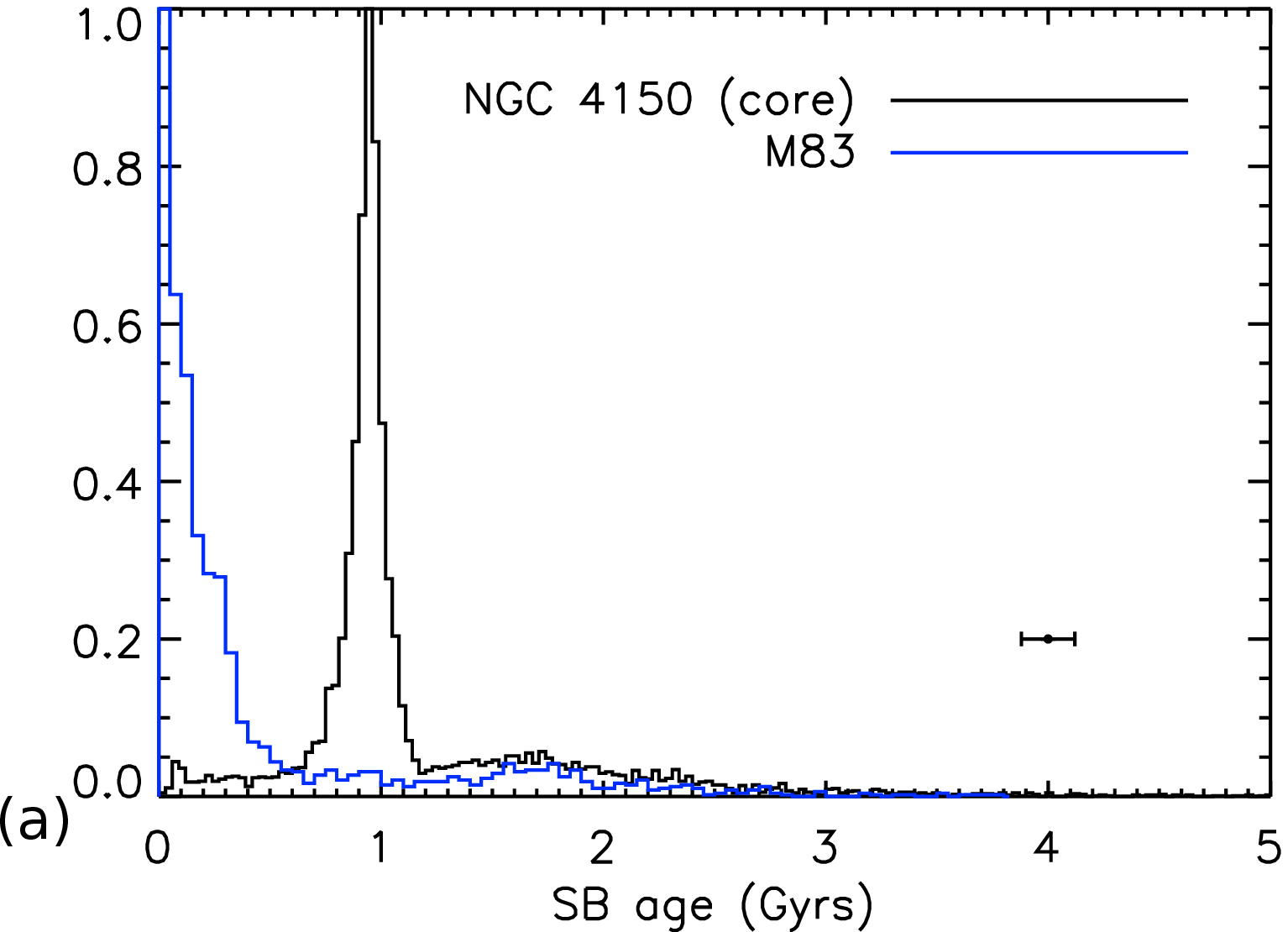}&
\includegraphics[width=75mm]{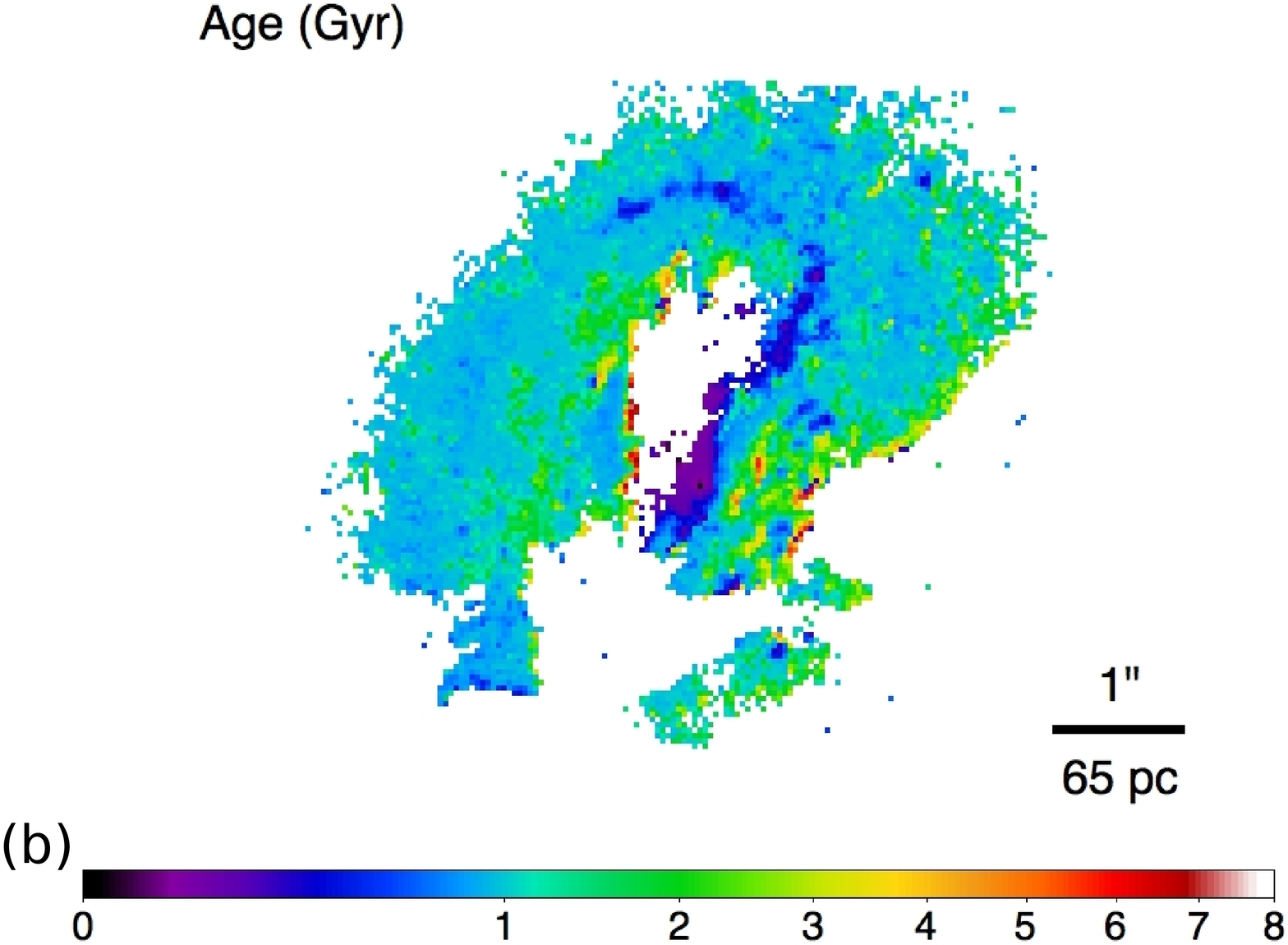}\\
\includegraphics[width=70mm]{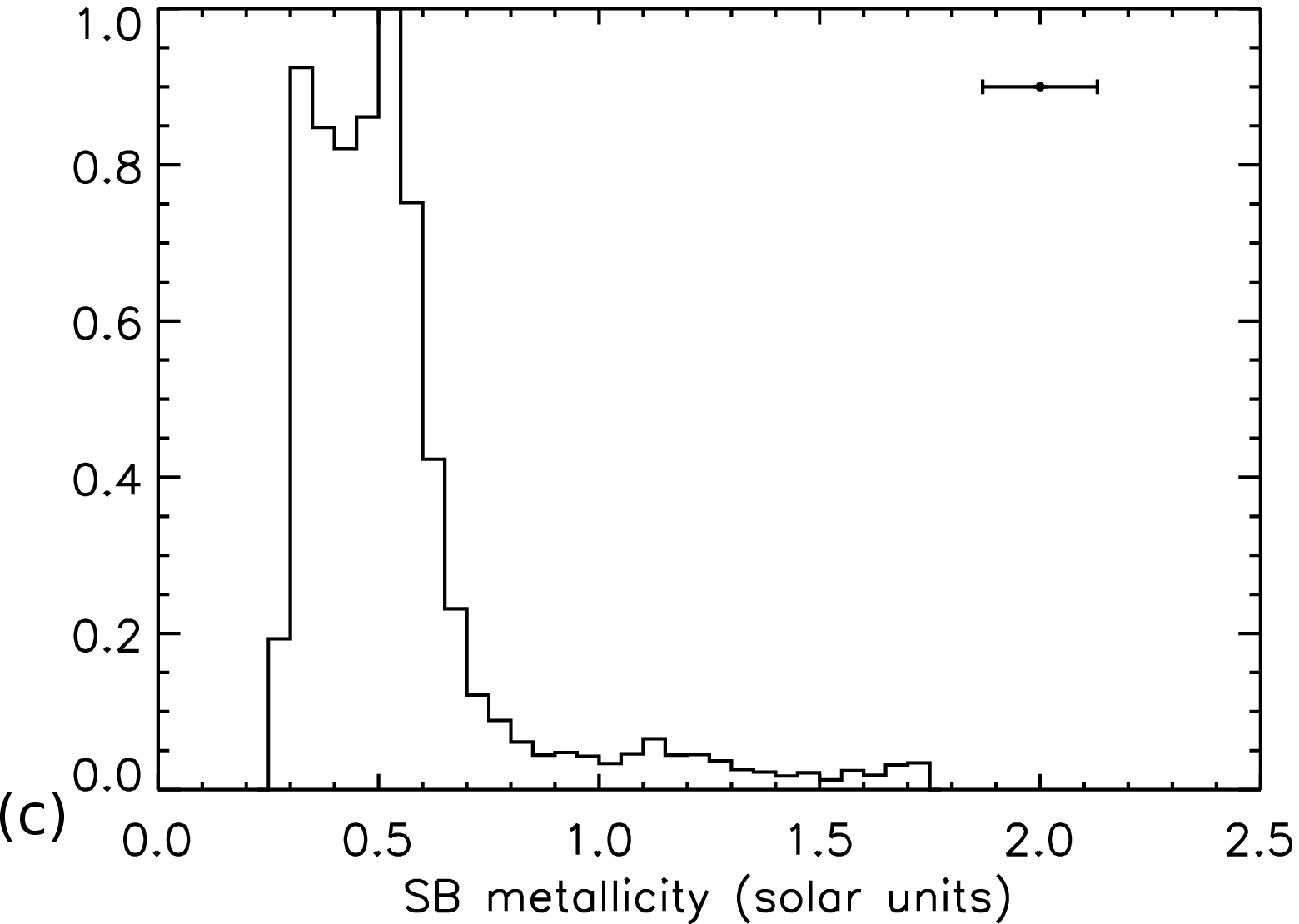}&
\includegraphics[width=75mm]{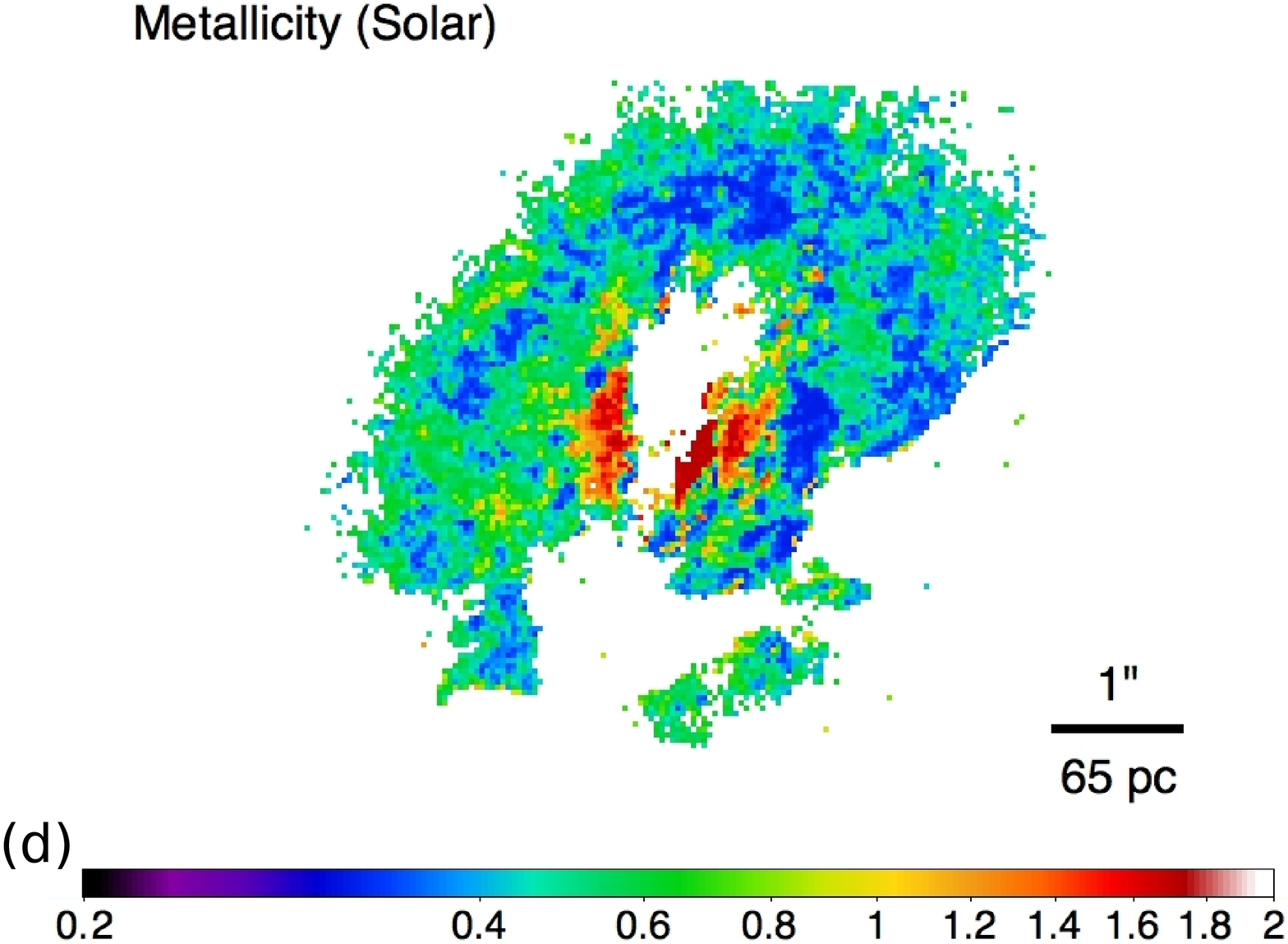}\\
\includegraphics[width=70mm]{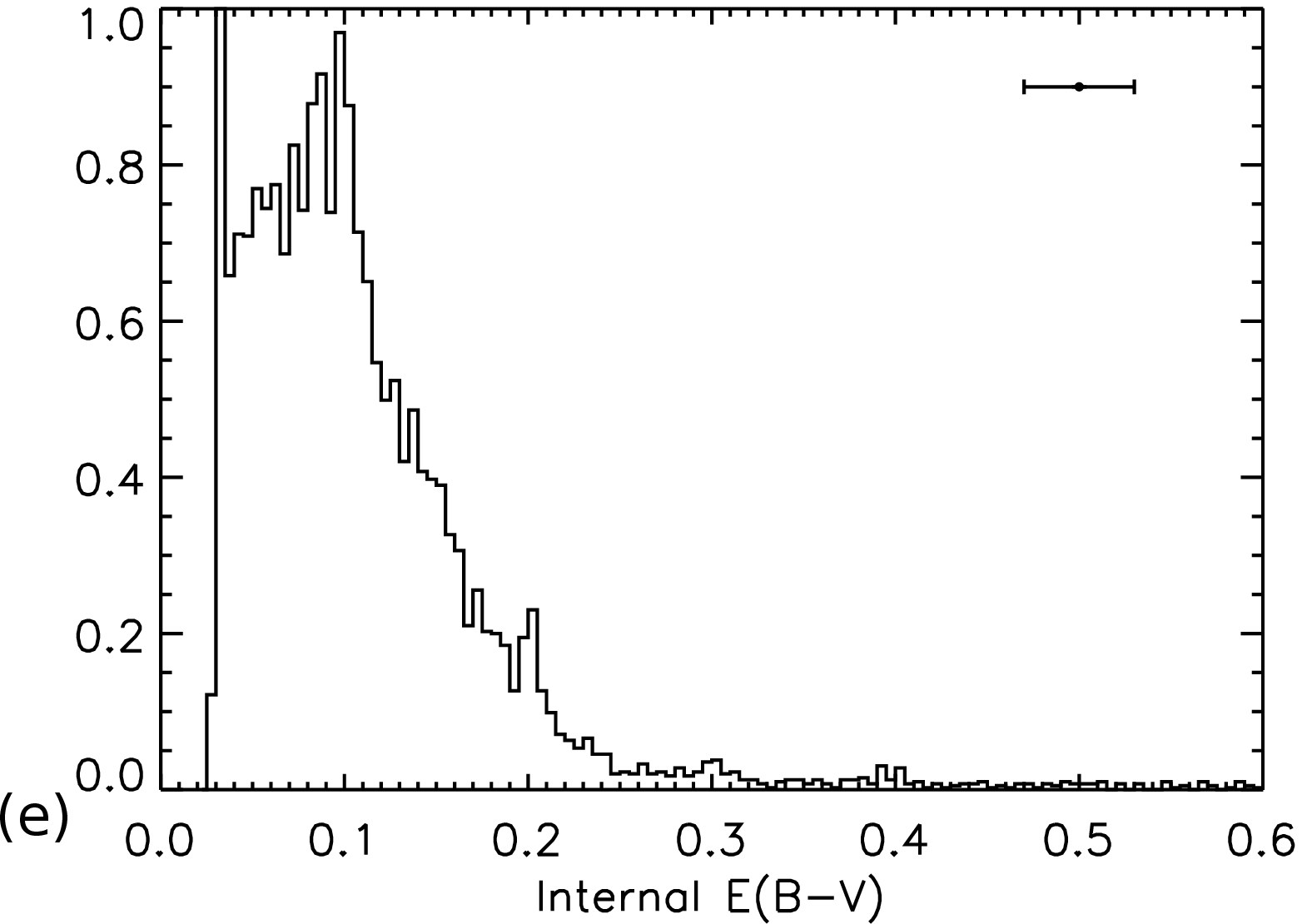}&
\includegraphics[width=75mm]{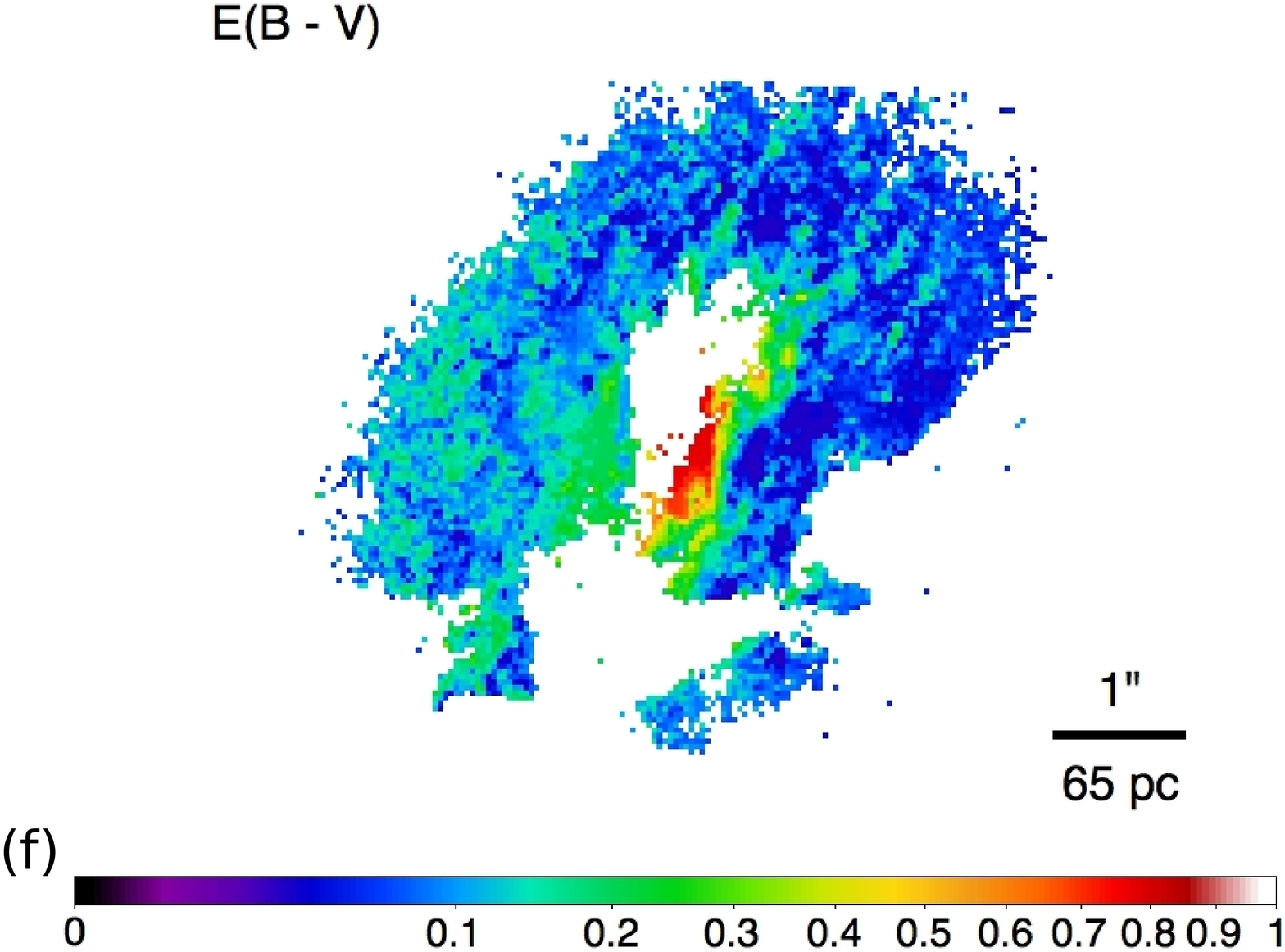}
\end{tabular}
\caption{Results of the parameter fitting routine shown as 1D pixel distributions and 2D maps.  Fitting was performed on all pixels with signal-to-noise greater than 5 in all five broadband filters.  The spatial extent of our parameter fitting was therefore limited by the depth of the $NUV$ observations (see Fig.~\ref{fig:pix_phot}a).  The error bars on the histogram plots indicate the typical 1$\sigma$ errors associated with the fitted parameters.   {\bf (a\&b)}: Age of the most recent burst of star-formation in the central pixels of NGC 4150.  %The histogram plot shows a narrow peak at ~0.9 Gyr (black line).  This is compared with the age distribution of pixels from a currently star forming region in M83, which shows a clear peak at age zero (blue line).
{\bf (c\&d)}: Metallicity of the most recent starburst in NGC 4150.  %This shows a peak at sub-solar metallicity, with a median value of ~0.5 solar and truncated at 0.3 solar.  Since the youngest stars are most likely to have formed from the gas accreted during the merger event, we can reasonably assume that the truncated peak is representative of the gas-phase metallcity of the galaxy that merged with NGC 4150 roughly 1 Gyr ago.  From the mass-metallicity relation of Tremonti et al. (2004) we estimate the mass of the accreted galaxy to be $\sim$ 4 x 10$^{8}$\msol, roughly 1/15 the mass of NGC 4150.
{\bf (e\&f)}: Extinction values for each of our fitted pixels.  Note that the pixel-by-pixel photometry was not corrected for Galactic extinction (E(B-V) = 0.018; Schlegel et al. 1998) prior to running the parameter estimation.  Furthermore, the 1D distribution is biased towards lower values of reddening, since, by setting a S/N limit of 5, we ignore pixels with high extinction.  See text in \S~\ref{sec:results} for a full discussion of the parameter distributions and maps.%***When we have parameter maps will require comments on position of youngest, most metal poor/rich and most extinguished stars.  May be necessary to split age, metallicity and extinction into separate figures if the caption becomes too unwieldy***
}
\label{fig:param_results}
\end{figure*}

\begin{figure*}
\centering
\begin{tabular}{cc}
\includegraphics[width=70mm]{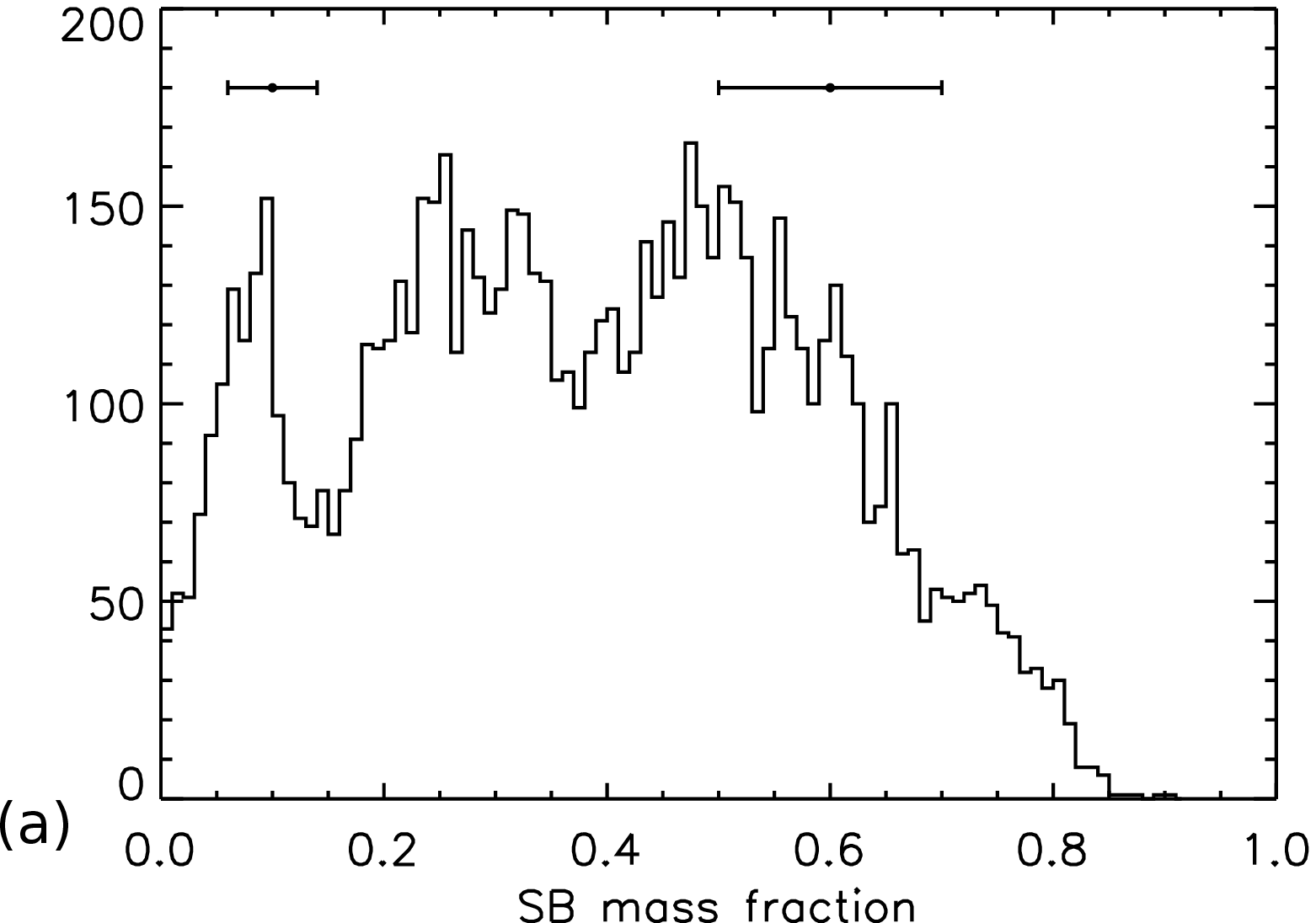}&
\includegraphics[width=75mm]{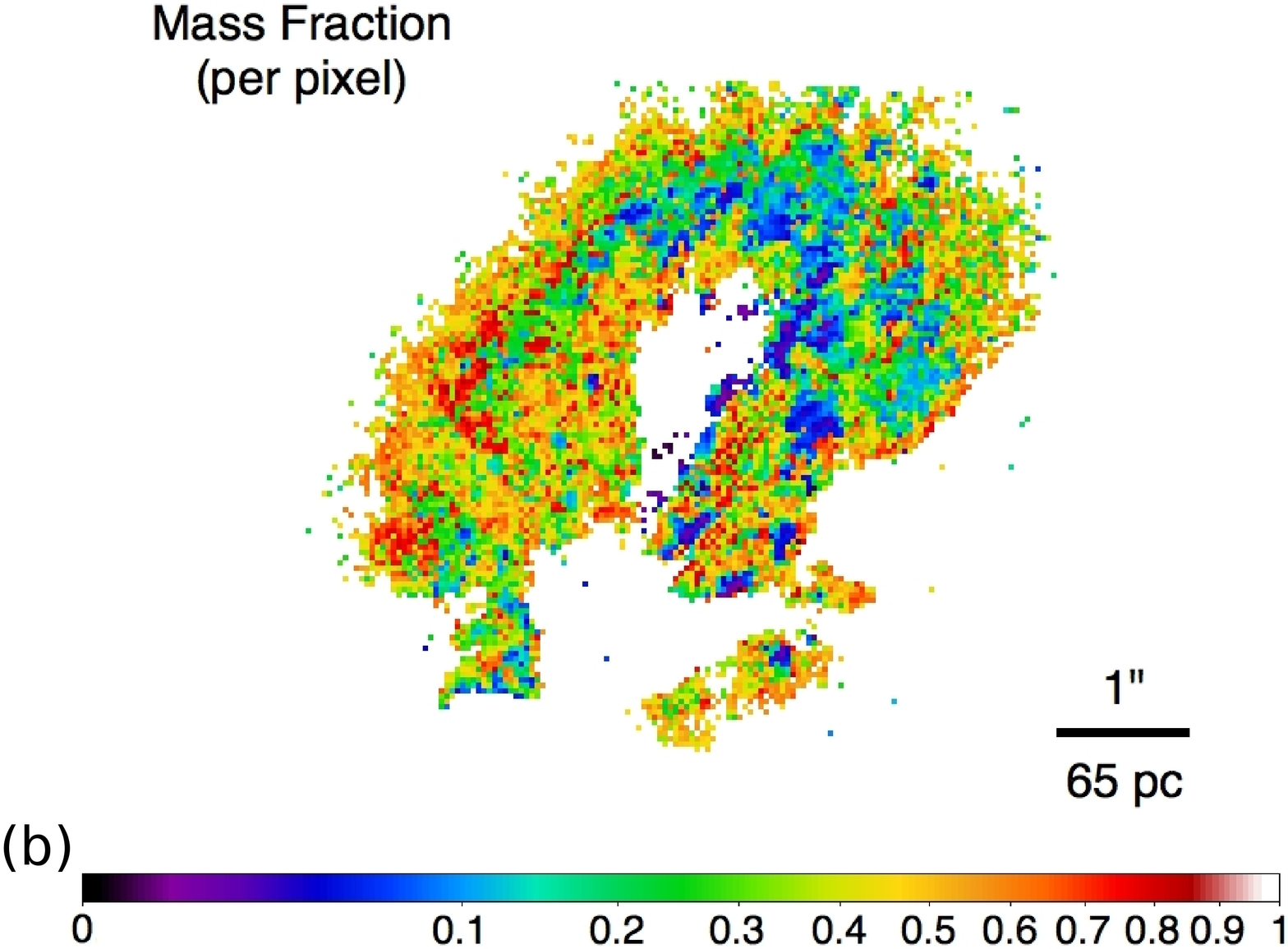}
\end{tabular}
\caption{{\bf (a)}: 1D distribution and {\bf (b)}: 2D map of stellar mass fraction (per pixel) of the most recent burst of star-formation in the central pixels of NGC 4150.  The two error bars on the histogram plot indicate the typical 1$\sigma$ uncertainties on fitted mass fractions of less than 0.2 (small error bar) and greater than 0.2 (large error bar).  See text for details.}
\label{fig:mass_fraction}
\end{figure*}

\begin{figure*}
\centering
\begin{tabular}{cc}
\includegraphics[width=80mm]{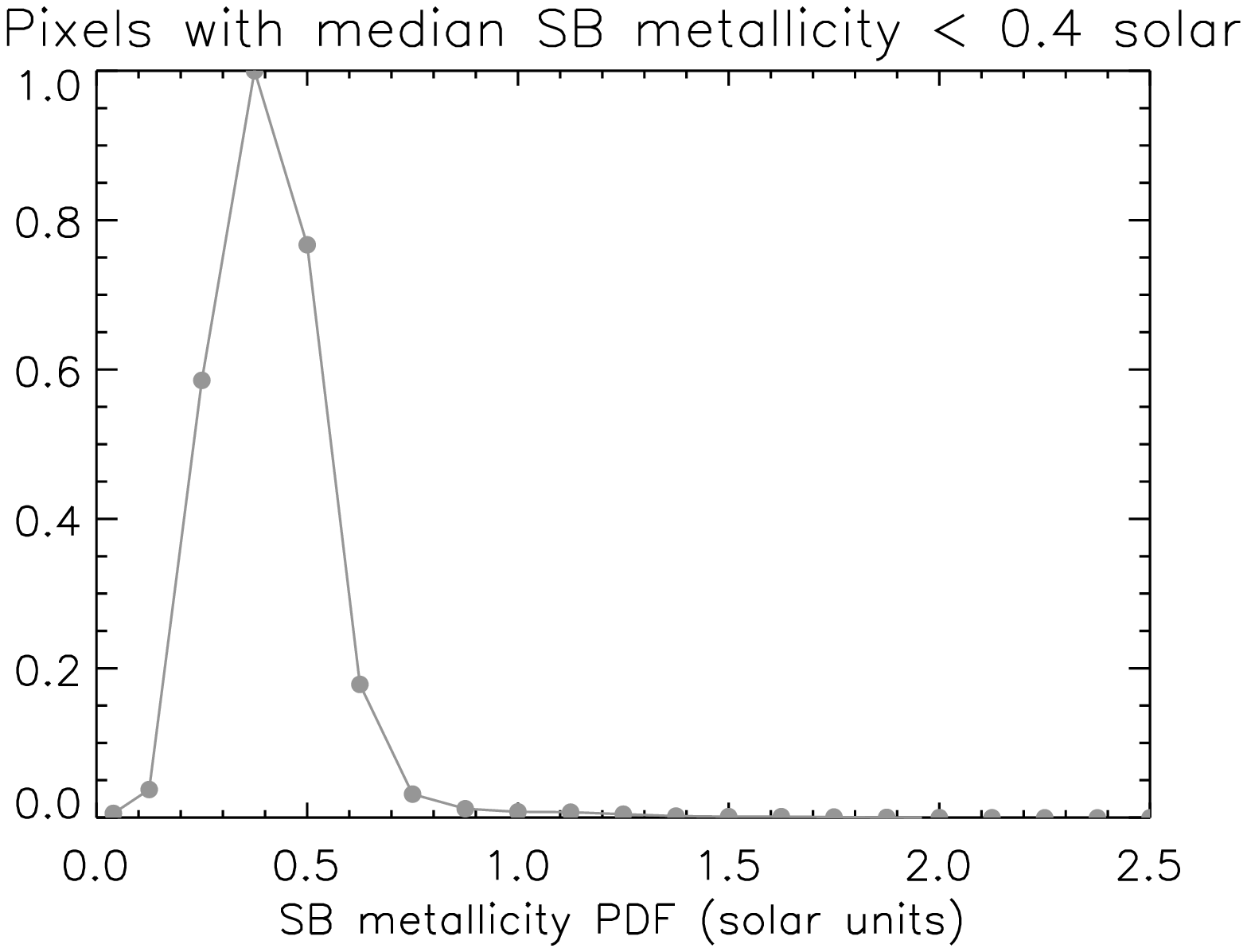}
\includegraphics[width=80mm]{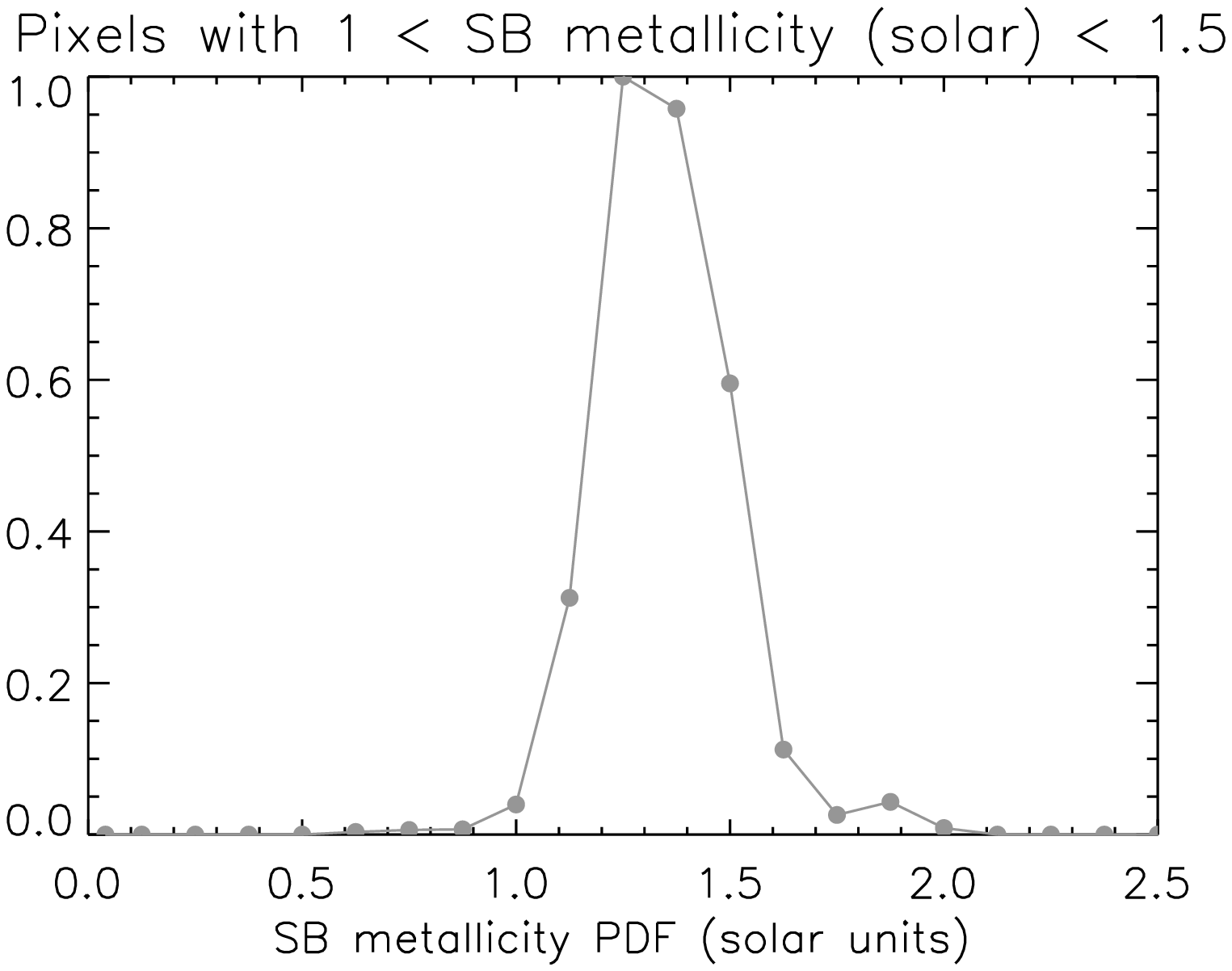}
\end{tabular}
\caption{The combined, marginalized PDFs for pixels with SB (star burst - i.e. RSF) metallicities less than 0.4 Z$_{\odot}$ (left panel) and RSF metallicities between 1 and 1.5 Z$_{\odot}$ (right panel).  It is apparent that the low and high metallicity populations are indeed distinct.  The separation of these two populations is made possible due to the inclusion of UV data, which leads to a reduction in the degeneracies between age, metallicity and extinction.}
\label{fig:metal_pdf}
\end{figure*}

We have parametrised the properties of the RSF (age, mass fraction, metallicity and internal dust content - Figs.~\ref{fig:param_results}\,\&\,\ref{fig:mass_fraction}) in the NGC 4150 pixels by comparing the observed $NUV$-optical photometry in five filters ($F225W, F336W, F438W, F555W, F814W$) to stellar models. The typical age of the RSF is around 0.9 Gyrs, consistent with the similarity of the $NUV$-optical colours to post-starburst systems.  The RSF age map reveals somewhat younger ($\sim$50 - 500 Myr) substructure, including a {\em clumpy} stream of stars spiraling into the galaxy core.  %\textcolor{red}{We suggest that this feature is a tidal tail left behind by the purported galaxy merger.  **Not sure we can say this.**}

We found the typical RSF metallicity - which is representative of the metallicity of the gas fuelling star formation - to be $\sim$0.3 - 0.5 Z$_{\odot}$, but note that within 0.75$\arcsec$ of the galaxy centre the recently formed stars have metallicities in the range 1.0 to 1.7 Z$_{\odot}$.  We propose that the most central RSF may have been fuelled by {\em metal-rich} gas already present in NGC 4150 that was caused to fall into the galaxy centre as a result of the merger, while the surrounding stellar population formed from {\em metal-poor} gas accreted during the merger.  Assuming this scenario to be correct, the lowest RSF metallicities (0.3 - 0.5 Z$_{\odot}$) plausibly indicate the gas-phase metallicity (GPM) of the accreted galaxy.  Comparison to the local mass-GPM relation \citep{2004ApJ...613..898T} suggests (crudely) that the mass of the accreted system is $\sim3\times10^8$ M$_{\odot}$, making NGC 4150 a 1:20 minor merger.  Summing the error-weighted mass fractions of RSF in each of the pixels, we found that the RSF contributes approximately 2-3 percent of the total stellar mass of the galaxy.

%Since we have suggested that NGC 4150 experienced a galaxy merger in the recent past it is worth commenting on the presence or otherwise of satellite galaxies.  Although associated with the Coma I cloud, NGC 4150 is located in a very low density environment \citep{2006MNRAS.371..157M}, with no obvious companion galaxies visible in deep survey observations (e.g. SDSS).  Several galaxies, of much smaller angular size than NGC 4150, are observed along lines-of-sight projected in and around the galaxy disk, but since none show evidence of gravitational disruption or tidal bridges connected to NGC 4150, all are identified as distant, background galaxies.  In our preferred scenario we have proposed that NGC 4150 merged with a galaxy approximately 1/20 of its mass, around 1 Gyr ago.  We estimate such a galaxy to have had an apparent $B$-band magnitude of $\sim$15-16 (scaled from NGC 4150 $B$-band magnitude), and we would certainly be able to detect similar mass galaxies if they existed around NGC 4150 today.  However, the apparent isolation of NGC 4150 does not rule out the merger scenario.  For example, it may already have merged with all satellite galaxies that might otherwise have been detected in deep survey observations.  Alternatively, peculiar velocities within the Coma I cloud may have caused NGC 4150 to move a significant distance from once nearby companions during the last Gyr.  Ultimately, the combined body of chemical, kinematical and morphological evidence strongly supports a recent minor merger as the trigger for the RSF in NGC 4150.

This work reaffirms our hypothesis that minor mergers play a significant role in the evolution of ETGs at late epochs \citep[e.g. see recent papers by][]{2010arXiv1001.2141K,2010ApJ...710.1170L}.  While many previous works have relied upon integrated photometry of large samples of ETGs, this WFC3 study confirms the value of 2D analyses of individual, nearby systems in which the structure and morphology of the constituent stellar populations can be resolved.  In a forthcoming paper (Kaviraj et al. in prep.) we will present a study of the globular cluster (GC) population in NGC 4150 using our WFC3 data, while future HST observations will lead to similar investigations of several other nearby, early-type galaxies.

\acknowledgements

We thank an anonymous referee for helpful comments and suggestions.  This paper is based on Early Release Science observations made by the WFC3 Scientific Oversight Committee.  We are grateful to the Director of the Space Telescope Science Institute for awarding Director's Discretionary time for this program.  Finally, we are deeply indebted to the brave astronauts of STS-125 for rejuvenating HST.  Support for Program numbers 11359/60 as provided by NASA through a grant from the Space Telescope Science Institute, which is operated by the Association of Universities for Research in Astronomy, Incorporated, under NASA contract NAS5-26555.  RMC acknowledges funding from STFC through research grant DBRPDV0.  SK acknowledges a Research Fellowship from the Royal Commission for the Exhibition of 1851, an Imperial College Research Fellowship, a Senior Research Fellowship from Worcester College Oxford and support from the BIPAC Institute at the University of Oxford.  SKY was supported by the Korean government through the Korea Research Foundation Grant (KRF-C00156) and theÊKorea Science and Engineering FoundationÊgrant (No.Ê20090078756).

\bibliographystyle{aa}
\bibliography{wfc3bibtex}

\newpage

\begin{table}
\caption{Properties of NGC 4150}
\begin{center}
\begin{tabular}{llr}
\hline\hline
$\alpha_{J2000}$ & $12^{h}10^{m}33^{s}.67$ & 1 \\
$\delta_{J2000}$ & $+30\degr24\arcmin05\arcsec.9$ & 1 \\
Morphological type & ${\rm SA0^0(r)?}$ & 2\\
Position angle & $148\degr$ & 1\\
Inclination angle & $58\degr$ & 1\\
${\rm v_{Heliocentric}}$ & $219\pm18$ ${\rm km\,s^{-1}}$ & 1\\
$\mu$ &  $30.66\pm0.16$ mag & 3\\
Galactic reddening & E(B - V) = 0.018 mag & 4\\
$M_{B}$ & -18.48 mag & 5\\
$(B - V)_{e}$ & 0.83 mag & 1\\
${\rm Log}\,L_{B}$ ($L_{B,\odot}$) & 9.50 & 6\\
${\rm Log}\,L_{Ks}$ ($L_{Ks,\odot}$) & 10.02 & 6\\
$M_{*}$ & $6.3^{+3.1}_{-2.1}\times10^{9}$ $M_{\odot}$ & 7\\
$M_{\rm H_2}$ & $3.8 - 6.6\times10^{7}$ $M_{\odot}$ & 8, 9\\
%SFR$_{24\,\mu {\rm m}}$ & $4.0\times10^{-2}$ $M_{\odot}\,{\rm yr^{-1}}$ & 6\\
\hline\hline
\end{tabular}
\end{center}
\footnotesize{(1) LEDA (http://leda.univ-lyon1.fr/); (2) NED (http://nedwww.ipac.caltech.edu/); (3) Distance modulus is mean of \citet{2003ApJ...583..712J} and \citet{2005MNRAS.361..330R} SBF measurements; (4) From \citet{1998ApJ...500..525S}; (5) Absolute $B$ magnitude from \citet{2002MNRAS.329..513D}; (6) From \citet{2009ApJ...695....1T}; (7) NGC 4150 stellar mass calculated using $M/L$ - colour relations from \citet{2003ApJS..149..289B} and assuming a \citet{1993MNRAS.262..545K} IMF (NGC 4150: $M_{*}/L_{Ks}$ =  $0.6^{+0.3}_{-0.2}$); (8) Molecular hydrogen masses from \citet{2003ApJ...584..260W} and (9) \citet{2007MNRAS.377.1795C}}
\label{tab:ngc4150_props}
\end{table}

\begin{table}
\caption{HST WFC3/UVIS observations of NGC 4150}
\begin{center}
\begin{tabular}{llc}
\hline\hline
Date & Filter & Exposure Time\\
& & (s) \\
 \hline
2009 Oct 30 & F225W & 3252\\
2009 Nov 09 & F336W & 2486\\
2009 Nov 09 & F438W & 2173\\
2009 Nov 09 & F555W & 1414\\
2009 Oct 30 & F657N (H$\alpha$ + [N II]) & 2508\\
2009 Oct 30 & F814W & 2048\\
\hline\hline
\end{tabular}
\end{center}
\label{tab:obs_table}
\end{table}

%\begin{figure*}
%\centering
%\begin{tabular}{ccc}
%\includegraphics[width=35mm]{NUV-V_5sig_labelled.eps}&
%\includegraphics[width=60mm]{n4150_pixcol_dist.ps}&
%\includegraphics[width=60mm]{n4150_col_col.ps}
%\end{tabular}
%\caption{{\bf (a)}: NUV - V colour map of central 6$\arcsec$ x 6$\arcsec$ of NGC 4150 using a signal-to-noise cut of 5$\sigma$.  {\bf (b\&c)}: Colour distribution and colour-colour plots of pixels in the central region of NGC 4150 (red) compared with pixel colours from a star-forming region in M83 (blue - also imaged with WFC3) and integrated colours of nearby ETGs \citep[black;][]{2007ApJS..173..619K}.  Also plotted in the colour-colour diagram are integrated colours of nearby E+A galaxies \citep[green;][]{2007MNRAS.382..960K}, post-starburst, major-merger remnants, with high mass-fractions of recent ($<$ 1 Gyr) star-formation.  The central colours of NGC 4150 are consistent with a post-starburst stellar population, falling between the currently star-forming M83 and the old, passively-evolving ETGs.  Redder optical colours (F438W-F555W) than E+A galaxies suggest a lower mass-fraction of recent star-formation in NGC 4150.}
%\label{fig:pix_phot}
%\end{figure*}

\label{lastpage}
 \end{document}